\documentstyle[eqsecnum,aps,prb]{revtex}

\begin{document}
\draft

\title{Capillary Wave Scattering From  a Surfactant
Domain}

\author{T. Chou}
\address{Department of Physics, Harvard University,
Cambridge, MA 02138}

\author{S. K. Lucas and H. A. Stone}
\address{Division of Applied Sciences, Harvard University,
Cambridge, MA 02138}

\date{\today}
\maketitle

\begin{abstract}
The study of capillary wave scattering by a circular region
with different interfacial properties from the rest of an
otherwise homogeneous interface is motivated by
experiments on wave attenuation at a monolayer covered
air-water interface where domains of one surface phase
are dispersed in a second surface phase. Here the
scattering function is calculated for an incident wave of
frequency $\omega$ (wavevector $k_0$) scattering from
an isolated circular domain of radius $a$ with surface
tension $\sigma_{1}$ which is imbedded in an otherwise
infinite interface of surface tension $\sigma_{0}$. The
underlying fluid is treated as irrotational and the
three-dimensional flow problem coupling the
heterogeneous surface to the underlying liquid is
reduced to a set of dual integral equations, which are
solved numerically.  With this solution the scattering
amplitudes and the total scattering cross sections are
calculated as a function of the surface tension ratio
$\sigma_{0}/\sigma_{1}$ and incident wavenumber
$k_0 a$.  The analogous problem of a discontinuous
change in bending rigidity is also considered and the
solution to the complete viscous problem is outlined in
an appendix.  Experimental implications of these results
are discussed.
\end{abstract}

\section{Introduction}

The interfacial region separating two distinct fluids has
significant dynamical effects on many transport
processes. In particular, fluctuations in the interface
shape are present in problems as diverse as classical
propagation of capillary-gravity waves, fluctuations of
cell membranes, and mass transport across the ocean
surface.  The majority of analyses of such surface waves
have treated a fluctuating interface with homogeneous
properties.  Nevertheless, due to the typical
phase-separated structure of monolayer covered
interfaces or the presence of impurities, the surface may
in fact be heterogeneous.\cite{REV,REV2} For example,
interfacial tension or bending rigidity, two material
properties used to characterize the interfacial resistance
to deformation, may vary along an interface. In this paper,
we analyze the effect of a circular domain of different
interfacial properties on the propagation of a capillary
wave and so deduce the scattering characteristics.

A typical feature of surface fluctuations with wavelength
$\lambda$ is that the velocity in the surrounding fluid
may be disturbed to a distance $\lambda$ with a
magnitude proportional to the magnitude of the restoring
force. Hence, in the neighborhood of a substantial change
in surface properties, it is likely that local transport
processes are affected.  A first approach to
understanding the dynamics of a heterogeneous interface
is to describe the mechanism of a plane wave scattering
from an isolated surface inhomogeneity.\cite{NOSKOV}
However, almost all previous experimental and
theoretical research on surface wave scattering has
focused on the effects of solid bodies in contact with the
bulk fluid; \cite{DEAN} e.g. scattering produced by
nonuniform bottoms \cite{DDS,BELZONS}, floating
objects\cite{MEI,JOHN}, and solid structures such as
cylinders protruding from the interface.\cite{MEI,MH}
These studies have concentrated on the gravity wave limit
with a few also treating capillary wave propagation.
There is a rich mathematical literature describing gravity
wave scattering by floating surface objects (such as a
circular disk) with papers dating back 40 years.
\cite{JOHN}

There are now many experimental observations,
including surfactant systems and lipid monolayers, of the
complex, heterogeneous surface structures that develop
when amphiphilic molecules reside at an air-water
interface.\cite{REV,REV2}  Frequently, the interface
consists of circular domains (with radii tens of microns
ranging up to a  centimeter) of one phase interspersed in
a second phase with different surface concentration.  The
surfactant concentration differences locally change the
parameters which govern the stress balance at the
interface. Both the surface tension and bending rigidity of
the interface can thus be altered and in these small
mesoscopic systems, it is surface tension, or bending
rigidity, rather than gravity, that provides the dominant
restoring force to surface deformations.  Therefore, we
study capillary wave propagation in the presence of an
isolated circular island with different surface properties
from the surrounding interface. Our work provides an
example of wave scattering due to a {\it boundary
condition} heterogeneity.

Experiments on capillary wave propagation at fluid-gas
interfaces covered with surfactants can show a dramatic
increase in the damping rate of the surface capillary
waves.\cite{MIY,WANG,LUC,EXP1,EXP2,EXP3} In these
experiments, a wave maker generates a capillary wave
and either the dispersion relation, wave amplitude, or the
width of the spectral peaks generated from light
scattering of a thermally fluctuating interface, is
measured, and the damping coefficient characterizing the
attenuation of the interface deformation is extracted.  The
damping coefficient is observed to depend on the surface
coverage and appears to have a maximum near the
regime where surface phases
coexist.\cite{LUC,EXP1,EXP2,EXP3} The damping rate may
also depend on the typical domain sizes which coexist in
the monolayer.\cite{WANG} Since the damping rates may
be five times greater than that due to viscous damping, a
mechanism including the effects of wave scattering by
surface domains has been suggested,\cite{NOSKOV,TC2}
but there is as yet no proven explanation for the
substantial damping rate increase.

Boundary conditions for gravity waves do not involve
higher mixed derivatives and as gravity acts uniformly
along the interface, surface wave scattering may be
produced by imposed boundary conditions, {\it e.g.} a
fixed dock occupying a half-plane on an infinite free
surface.  Such problems  have been studied with
Wiener-Hopf techniques.\cite{HEINS} Furthermore, wave
scattering from an {\it intrinsic} boundary condition
heterogeneity can occur when a varying interfacial
tension or rigidity is present. In this case, the higher
mixed derivatives that occur in the boundary conditions
couple to the nonuniform surface parameters to produce
surface wave scattering. This type of problem has
received little attention. In particular, approaches to this
latter problem in the past have been
approximate,\cite{NOSKOV} or treated a specific one
dimensional geometry with additional assumptions
concerning the stability of the problem.\cite{GOU}

Here, we study analytically surface wave scattering from a
circular discontinuity in surface tension or bending
rigidity.  In Section II, we outline the boundary value
problem for an irrotational subphase.  Hankel transforms
may be used to describe the velocity potential and
application of the normal stress balance at the
heterogeneous interface leads to a system of dual integral
equations with Bessel function kernels.  This system is
solved numerically. The analogous problem accounting
for viscous effects is described in an appendix.  For a
benchmark comparison, we also derive results for
scattering from a stretched, free membrane with an
imbedded region of different tension.  In Section III, we
consider capillary wave scattering from a circular region
of different interfacial bending rigidity and compare the
results to bending wave scattering along a membrane
with a circular inclusion of different flexural rigidity. The
results of these different models are discussed and
compared in Section IV. A discussion of the experimental
applicability of these two problems, the criteria for
neglecting viscous effects, and the implications for
capillary wave damping are also given.

\section{Capillary Waves and Surface Tension
Discontinuities}

\subsection{Formulation}

Consider the propagation of small amplitude capillary
waves on a fluid with kinematic viscosity $\nu$, density
$\rho$ and of depth $H$ (Figure 1a). A plane wave from
far away impinges on a circular domain of radius $a$
which resides on the otherwise homogeneous interface.
First, we treat the classical capillary wave problem with
surface tension as the restoring force.  Second, we study
scattering from a bending rigidity discontinuity in
Section III. In all cases, gravitational effects on wave
propagation are neglected.

For a surface tension discontinuity, the constant surface
tension of the interface exterior to the domain is denoted
$\sigma_{0}$ and of the domain is denoted
$\sigma_{1}.$ Mechanical stability of such a circular
domain requires that a line tension act along the domain
boundary.  In monomolecular films, the line tension arises
from the van der Waals interactions between the
molecules.\cite{REV,REV2} For the dynamical calculation
reported here, only effects of surface tension are
important and line tension effects are higher order since
the domain is assumed to deviate only slightly from a
circular shape.  We thus apply linearized capillary wave
theory to obtain the scattering produced by the
discontinuous region of surface tension.  Both the velocity
field and height of the fluctuating interface are
calculated.  This interface displacement is often measured
experimentally by light scattering
techniques.\cite{EXP1,EXP2,EXP3}

Assuming an irrotational flow,
the velocity may be represented by a potential

\begin{equation}
v_{i}(x_{j}, t) \equiv {\partial \phi(x_{j},t)
\over \partial x_{i}}.
\end{equation}

\noindent The flow is assumed to be incompressible
so  that

\begin{equation}
\nabla^{2}\phi = 0.\label{LAPLACE}
\end{equation}

\noindent It is convenient to use cylindrical polar
coordinates ($R,\theta,Z$) and later the dimensionless
coordinates ($r,\theta,z$) are used.  For small
amplitude disturbances, the momentum equation is
linearized, so that the pressure $p = -\rho
\partial_{t}\phi$, and the boundary conditions are also
linearized and applied at $Z=0.$ Finally, the surface
height is denoted $\eta(R,\theta, t)$.  Effects of viscosity
are treated in Appendix A.

The boundary conditions to be applied are a kinematic
constraint

\begin{equation}
\begin{array}{ll}
\displaystyle
{\partial \eta \over \partial t} = {\partial
\phi\over \partial Z} & \quad \hbox{at}
\quad Z=0,\label{KIN}
\end{array}
\end{equation}

\noindent a dynamic condition which follows from the
normal stress balance,

\begin{equation}
\begin{array}{ll}
\displaystyle
\sigma\nabla_{\perp}^{2}\eta = \rho {\partial \phi
\over \partial t} &\quad \hbox{at}\quad Z=0, \label{BC0}
\end{array}
\end{equation}

\noindent where $\nabla_{\perp}^{2}$ is the
two-dimensional Laplacian expressed in the surface
coordinates, and an impenetrability condition at the
bottom of the trough

\begin{equation}
\begin{array}{ll}
\displaystyle {\partial \phi\over \partial Z} = 0 &
\quad \hbox{at} \quad Z=-H.
\label{BOTTOM}
\end{array}
\end{equation}

\noindent Taking a time derivative of (\ref{BC0}) and
using the kinematic condition (\ref{KIN}) yields

\begin{equation}
[\sigma\nabla_{\perp}^{2}\partial_{Z}
-\rho\partial_{t}^{2}]\phi= 0,
\label{BCN}
\end{equation}

\noindent which, along with (\ref{LAPLACE}) and
(\ref{BOTTOM}), govern the evolution of the velocity
potential.

For a plane wave propagating in a fluid with
homogeneous surface properties and depth $H$,  $\phi
\propto e^{ik_{0}X-i\omega t}\cosh k(Z+H)$ determines
the well-known dispersion relation for surface capillary
waves excited at frequency $\omega$,

\begin{equation}
\rho\omega^{2} = \sigma k_{0}^{3} \tanh
k_{0}H,\label{DISP0}
\end{equation}

\noindent where $k_{0}= 2\pi/\lambda$ denotes the
wavevector and $\lambda$ is the wavelength. The  normal
stress contribution to (\ref{BC0}) due to the gravitational
acceleration $g$ may be neglected provided

\begin{equation}
k_{0} \gg \left({\rho g \over \sigma}\right)^{1/2}. \label{G}
\end{equation}

\noindent Thus, for the air-water interface, gravity can
neglected for wavelengths $\lambda = 2\pi/k \ll 1.7$ cm.

In most experimental situations, $\omega$ is a real, fixed
frequency determined by the wavemaker and  equation
(\ref{DISP0}) defines the wavevector $k_{0}(\omega)$ as a
function of $\omega$ for a particular $\sigma.$ When
the viscosity of the fluid is explicitly included in the
linear analysis of surface fluctuations, the wavevectors
have a positive imaginary component\cite{LEVICH}
(derived in Appendix A) corresponding to viscous
damping of the surface fluctuations.

\subsection{Scattering by a circular surface domain}

We now consider the disturbance to the plane progressive
wave caused by a circular domain on the interface.
This domain changes the surface tension within it such
that

\begin{equation}
\sigma(r) =
\sigma_{0}\Theta(R-a) + \sigma_{1}\Theta(-R+a)
\label{s0s1}
\end{equation}

\noindent where $\Theta(x)$ is the Heaviside function
which is unity for $x>0$ and is zero otherwise.

The velocity potential in the lower half-space can written
as an incident wave coming from $X \rightarrow -\infty$,
plus a scattered velocity potential, $\psi(R,\theta,Z)$,
which is assumed to be outgoing and to decay as
$R\rightarrow \infty$:

\begin{equation}
\phi(R,\theta,Z,t)
=  {\cosh k_{0}(H\!+\!Z)
\over \cosh\, k_{0}H}
e^{ik_{0}X-i\omega t}+ \psi(R,\theta,Z)e^{-i\omega t},
\label{SCATT}
\end{equation}

\noindent where $k_{0}$ is given by the dispersion
relation (\ref{DISP0}) with  $\sigma = \sigma_{0}$.  Using
the $\theta \rightarrow -\theta$ angular symmetry of the
problem, the scattered potential may be expanded as

\begin{equation}
\psi(R,\theta,Z) = \sum_{m=0}^{\infty}
\cos\,m\theta  \int_{0}^{\infty}
dk\, \psi_{m}(k) J_{m}(kR)
\left[{\cosh\, k(H\!+\!Z) \over \cosh\, kH}
\right],\label{DEF}
\end{equation}

\noindent where $\theta$ is measured from the incident
wavevector (see Figure 1a) and the $J_{m}(kR)$ terms
ensure that the potential remains bounded as
$R\rightarrow 0.$ Equation (\ref{DEF}) is a complete
expansion in Bessel functions and describes the analytic
function $\psi(R,\theta, Z)$ obeying the Laplace equation
in the domain $z \leq 0$ and satisfying the
no-penetration condition at $Z=-H$. Later in the analysis,
we will see that the choice of outgoing disturbances from
the domain (a radiation condition) requires that the first
kind Hankel functions, $H_{m}^{(1)}(kR)$, represent the
dominant contribution for large $R$.

The functions $\psi_{m}(k)$ are determined by first
substituting (\ref{SCATT}) into (\ref{BCN}). Since the
incident wave in (\ref{SCATT}) automatically satisfies the
normal stress boundary condition at the interface, then
outside the circular region ($R>a$), where the surface
tension is $\sigma_{0}$, we have

\begin{equation}
\begin{array}{ll}
\quad [\sigma_{0}\nabla^{2}_{\perp} \partial_{Z} +
\rho\omega^{2}]\psi(R,\theta,Z) \vert_{Z=0} = 0
& \quad\quad \quad \quad \quad (R>a)  , \label{NO1}
\end{array}
\end{equation}

\noindent while inside the circular domain we have

\begin{equation}
\begin{array}{ll}
[\sigma_{1}\nabla^{2}_{\perp}\partial_{Z} + \rho
\omega^{2}]\psi(R,\theta,Z)\vert_{Z=0} =
k_{0}^{3}(\sigma_{1} - \sigma_{0})e^{ik_{0}R \cos \theta}
\tanh k_{0}H & \quad \quad (0\leq R < a).
\label{NI1} \end{array}
\end{equation}

\noindent We can further expand the right hand side of
(\ref{NI1}) using the identity \cite{GR}

\begin{equation}
e^{ik_{0}R cos \theta} =\left[ J_{0}(k_{0}R) +
2\sum_{p=1}^{\infty}i^{p}J_{p}(k_{0}R) \cos\,p\theta
\right].\label{EXPAND}
\end{equation}

Substituting (\ref{EXPAND}) and (\ref{DEF}) into (\ref{NI1})
and multiplying both sides of (\ref{NI1}) and (\ref{NO1})
by $\cos\,n\theta$, then integrating $\theta$
from $0$ to $2\pi$, we obtain a set of dual integral
equations for each $n\geq 0:$

\begin{equation}
\begin{array}{ll}
\displaystyle \int_{0}^{\infty}dk\, J_{n}(kR)
\psi_{n}(k)(\rho\omega^{2}-\sigma_{0}k^{3}
\tanh\,kH) = 0,  & \quad \quad \quad \quad (R>a) \\ [13pt]
\displaystyle \int_{0}^{\infty}dk \,J_{n}(kR) \psi_{n}(k)
(\rho\omega^{2}-\sigma_{1}k^{3}
\tanh\, kH ) =  \\
\hspace{1cm} \quad (2-\delta_{n,0})i^{n}
(\sigma_{1}-\sigma_{0})k_{0}^{3}
 J_{n}(k_{0}R) \tanh k_{0}H & \quad \quad \quad\quad
\quad (0\leq R<a).
\end{array}
\end{equation}

\noindent Rescaling $r = R/a, h = H/a, ka= q$, and letting
$\Lambda =\sigma_{0}/\sigma_{1}$ denote the
surface tension ratio, we can rewrite this
system of integral equations as

\begin{equation}
\begin{array}{ll}
\displaystyle \int_{0}^{\infty} F_{n}(q)J_{n}(qr)\,
dq = 0, & \quad\quad \quad\quad (r>1) \\[13pt]
\displaystyle \int_{0}^{\infty} F_{n}(q)G(q) J_{n}(qr)\,dq
=\Delta_{n} J_{n}(q_{0}r), &\quad
 \quad\quad \quad (0\leq r<1)
\end{array}\label{DIE2}
\end{equation}

\noindent where

\begin{equation}
F_{n}(q) \equiv
{(q_{0}^{3}\tanh\,q_{0}h-q^{3}\tanh qh)
\over i^{n} a} \psi_{n}(q/a),\label{Fq}
\end{equation}

\begin{equation}
G(q) \equiv {q^{3}\tanh qh -q_{1}^{3}
\tanh q_{1}h \over q^{3}\tanh qh-q_{0}^{3}
\tanh q_{0}h} , \label{Gq}
\end{equation}

\noindent and

\begin{equation}
\Delta_{n} \equiv (2-\delta_{n,0})(1-\Lambda)
q_{0}^{3}\tanh q_{0}h.\label{DELTA}
\end{equation}

\noindent In (\ref{Fq}), (\ref{Gq}), and (\ref{DELTA}), we
have introduced the dimensionless wavevectors

\begin{equation}
q_{j}^{3} \tanh q_{j}h\equiv
{\rho\omega^{2} a^{3} \over \sigma_{j}}
\quad \quad (j=0,1). \label{DISP}
\end{equation}

\noindent We note that $q_{1}=k_{1}a$ denotes the
dimensionless wavevector which would be measured if
the entire interface were homogeneous with surface
tension $\sigma_{1}.$

Finally, as mentioned above, the wavevectors can be
interpreted as having a small positive imaginary part
which ensures that the potentials vanish as
$R\rightarrow \infty.$  This behavior can be attributed to
the effects of small viscous damping.  The small damping
may be explicitly accounted for by studying the viscous
flow problem and taking the limit $\nu \rightarrow
0^{+}$. A dual integral equation formulation to the
viscous flow problem for finite $\nu$ is given for
completeness in Appendix A.  Here, we treat fully the
inviscid problem, but simply allow the wavevectors to
have an infinitesimally small positive imaginary part as is
common in scattering calculations.\cite{ARFKEN}

\subsection{Solution of the integral equations}

The flow problem contains three dimensionless
parameters: $q_{0}$, the  incident wavevector; $h$, the
fluid depth; and $\Lambda$, the surface tension ratio
characterizing the surface discontinuity. It is also clear
from (\ref{DELTA}) that $\Lambda=1$ gives
$\Delta_{n}=0$ for all $n\geq 0$ so that using
(\ref{DIE2}) and (\ref{Fq}) $\psi_{n} = 0$  and hence there
is no scattering.

The solution of the dual integral equations (\ref{DIE2})
for the unknown functions $F_n (q)$
can be readily determined numerically using a procedure
introduced by Tranter,\cite{TRANTER} which we describe
for completeness.  Begin by noticing the form of the
Weber-Schafheitlin discontinuous integral\cite{WATSON}

\begin{equation}
\int_{0}^{\infty}q^{1-\beta}
J_{2m+n+\beta}(q)J_{n}(qr)\, dq =\left\{ \begin{array}
{ll} \displaystyle {\Gamma(n+m+1)r^{n}
(1-r^{2})^{\beta-1} \over
2^{\beta-1}\Gamma(n+1)\Gamma(m+\beta)}
{\cal F}_{m}(\beta+n,n+1;r^{2}) & (0\leq r<1) \\
0 & (r>1)\label{WS}\end{array}  \right.
\end{equation}

\noindent where ${\cal F}_{m}(\beta+n,n+1;r^{2})$ is the
Jacobi polynomial of order $m$.\cite{GR}  Based upon the
analogy of the right-hand-side of (\ref{WS}) and
(\ref{DIE2}) for $r >1$, it is natural to expand the
unknown functions $F_{n}(q)$
in terms of Bessel functions as

\begin{equation}
F_{n}(q) \equiv q^{1-\beta}\sum_{m=0}^{\infty}
A_{m}^{(n)} J_{2m+n+\beta}(q),\label{Fn}
\end{equation}

\noindent where $\beta$ is a constant which may be
chosen later to improve the rates of convergence of the
required numerical integrations. With this representation
of $F_{n}(q)$, the first equation in (\ref{DIE2}) is
automatically satisfied.  Substituting (\ref{Fn}) into the
second equation in (\ref{DIE2}), we are left with

\begin{equation}
\begin{array}{ll}
\displaystyle \sum_{m=0}^{\infty}
A_{m}^{(n)} \int_{0}^{\infty}
dq\,q^{1-\beta}J_{2m+n+\beta}(q)J_{n}(qr)G(q)
= \Delta_{n} J_{n}(q_{0}r) & \quad \quad\quad 0\leq r <1.
\label{EQ2}
\end{array}
\end{equation}

\noindent Multiplying  both sides of (\ref{EQ2}) by
$r^{n+1}(1-r^{2})^{\beta-1}{\cal F}_{l}
(\beta+n,n+1;r^{2})$ and integrating $r$ from $0$ to $1$
gives for each $n$, where $n=0,1,2,3\ldots$,  a system of
equations

\begin{equation}
\sum_{m=0}^{\infty}A_{m}^{(n)} \int_{0}^{\infty} dq\,
q^{1-2\beta}J_{2m+n+\beta}(q)J_{2l+n+\beta}(q)
G(q) =q_{0}^{-\beta}
\Delta_{n}J_{2l+n+\beta}(q_{0}) \quad\quad l = 0, 1, \ldots \label{LIN0}
\end{equation}

\noindent Equation
(\ref{EQ2}) represents an infinite matrix  for the
coefficients $A_{m}^{(n)}$ and can be written succinctly
as

\begin{equation}
\begin{array}{ll}
\displaystyle \sum_{m=0}^{\infty} A_{m}^{(n)}
L_{ml}^{(n)}(\beta;q_{0},\Lambda) =
q_{0}^{2\beta-2}\Delta_{n}J_{2l+n+\beta}(q_{0})
& \quad \quad  (n,l \geq 0)
\label{LIN1}
\end{array}
\end{equation}

\noindent where

\begin{equation}
\begin{array}{l}
\displaystyle L_{ml}^{(n)}(\beta;q_{0},\Lambda)
= \int_{0}^{\infty}{dx \over x^{2\beta-1}}
{(x^{3}\tanh q_{0}hx -\Lambda \tanh q_{0}h)
\over (x^{3}\tanh q_{0}hx-\tanh q_{0}h)}
J_{2m+n+\beta}(q_{0}x) J_{2l+n+
\beta}(q_{0}x) \\
 \hspace{2.8cm}\displaystyle
\equiv \int_{0}^{\infty} dx I_{ml}^{(n)}
(q_{0},h,\Lambda;x) .
\label{L}
\end{array}
\end{equation}
\vspace{3mm}

\noindent The infinite set of linear equations (\ref{LIN1})
determines the coefficients of the expansion of the
functions $F_{n}(q)$ from which the Fourier
coefficients $\psi_{n}(k)$ are known by (\ref{Fq}) and
hence the scattered wave $\psi(r,\theta,z)$ is
determined.

The integral in equation (\ref{L}) has a pole on the real
axis at $x=1$ (i.e. $q = q_0$).  We may treat this pole in a
manner consistent with outgoing waves\cite{ARFKEN} by
interpreting the pole as having a positive imaginary part,
$Im\{q_{j}\} = \epsilon > 0$ and taking the limit
$\epsilon \rightarrow 0$.  This limit is consistent with
viscous effects becoming small, $\nu \rightarrow 0^{+}.$
In the limit $\epsilon \rightarrow 0$, the integral can be
evaluated by taking the path of integration through a
small semicircle below the pole at $x=1$ which  yields

\begin{equation}
\begin{array}{l}
\displaystyle L_{ml}^{(n)} =
i\pi(1-\Lambda)\left[{\sinh
q_{0}h \cosh q_{0}h \over  3\sinh q_{0}h \cosh
q_{0}h + q_{0}h}\right] J_{2m+n+\beta}(q_{0})
J_{2l+n+\beta}(q_{0})  \\[13pt]
\hspace{2.1cm} \displaystyle +\int_{0}^{1}dx
\left[I_{ml}^{(n)}(q_{0}, h,
\Lambda; x)+I_{ml}^{(n)}(q_{0}, h, \Lambda; 2-x)
\right] + \int_{2}^{\infty} dx\,
I_{ml}^{(n)}(q_{0}, h, \Lambda; x) \label{L2}\end{array}
\end{equation}

\noindent where we have separated the integrable real
part of the integral over $I_{ml}^{(n)}$ from the
contribution from the semicircle below the pole.  The
remaining integrals are evaluated numerically.

We next solve a truncated system of linear equations
(\ref{LIN1}) and so obtain the $A_{m}^{(n)}$'s which
depend on $q_{0}, h$, and $\Lambda$. The scattered
potential follows from

\begin{equation}
\psi(r,\theta,z;h) \! = \!\!
\sum_{n,m=0}^{\infty} \!\!
 i^{n}A_{m}^{(n)} \!\!
\int_{0}^{\infty}\!\! dq\,{q^{1-\beta}
\cos n \theta \over q_{0}^{3}
\tanh q_{0}h-q^{3}\tanh qh}{\cosh
q(h+z) \over \cosh qh}
J_{2m+n+\beta}(q) J_{n}(qr)
\label{psi}
\end{equation}

\noindent from which we find the interfacial height
displacement of the scattered part of the wave

\begin{equation}
\eta_{s}(r,\theta)
= {i \over \omega}{\partial \over \partial z}\psi
(r,\theta,z)\vert_{z=0}
= {i\over \omega} \!\!\sum_{n,m=0}^{\infty} \!\!
i^{n}A_{m}^{(n)} \!
\int_{0}^{\infty} \!\! dq\,{q^{2-\beta}\cos n\theta
\over q_{0}^{3}\tanh q_{0}h-q^{3}\tanh qh}
J_{2m+n+\beta}(q)J_{n}(qr).
\label{ETA}
\end{equation}

Equations (\ref{psi}) and (\ref{ETA}) also have poles on the
real axis at $q=q_{0}$, which we choose to integrate
using a semicircle contour below the pole.  This
treatment fixes the scattered waves as outgoing.  In
particular, to evaluate the integral in (\ref{psi}), we
replace the $J_{n}(qr)$ term  with ${1 \over
2}(H_{n}^{(1)}(qr)+H_{n}^{(2)}(qr))$.  For $r>1$,  the
integral over $J_{2m+n+\beta}(q)H_{n}^{(1)}(qr)$ is
closed by the contour in the upper half-plane shown in
Figure 2. The detour in quadrant one corresponds to a
branch cut at $q = \sqrt{i\omega a^2 /\nu}$ and is
derived in Appendix A.  Similarly, the
$J_{2m+n+\beta}(q)H_{n}^{(2)}(qr)$ integral is closed in
the lower half-plane.  Only the $H_{n}^{(1)}$ integral
encloses the pole at $q=q_{0}$.  Thus, only the outgoing
waves proportional  to $H_{n}^{(1)}(q_{0}r)$ are
important for large $r$.  The final result can be expressed
in the form

\begin{eqnarray}
\int_{0}^{\infty}dq\,{q^{1-\beta} \over
q_{0}^{3}\tanh
q_{0}h-q^{3}\tanh qh}{\cosh
q_{0}(h+z) \over \cosh
q_{0}h}J_{2m+n+\beta}(q)J_{n}
(qr) \simeq  \nonumber \\[13pt]
-\pi  i {\cosh q_{0}h \cosh q_{0}(h+z) \over
q_{0}^{\beta+1}(3\sinh
q_{0}h \cosh q_{0}h
+q_{0}h)}J_{2m+n+\beta}(q_{0})H_{n}^{(1)}
(q_{0}r)-\Gamma_{+}-\Gamma_{-} \quad (r>1)
\label{INT}
\end{eqnarray}
\vspace{2mm}

\noindent The contributions to the integral from the
integrations along the imaginary axis in Figure 2,
$\Gamma_{\pm}$, can be shown to be negligible
compared to $H_{n}^{(1)}(q_{0}r)$ as
$r\rightarrow \infty.$ Asymptotically,

\begin{equation}
H_{n}^{(1)}(q_{0}r) \simeq \sqrt{{2 \over \pi q_{0}r}}
e^{iq_{0}r-i\pi n/2-i\pi/4} \quad \quad\quad\quad
(r\rightarrow \infty)
\end{equation}

\noindent which leads us to write the far-field potential
in the form

\begin{equation}
\begin{array}{ll}
\displaystyle \psi(r,\theta,z;h) \simeq
f(\theta;\Lambda, h){e^{iq_{0}r} \over
\sqrt{r}}{\cosh q_{0}(h+z) \over \cosh q_{0}h}
& \quad \quad\quad (r \rightarrow \infty)
\end{array}
\end{equation}

\noindent where the scattering amplitude is obtained by
combining (\ref{psi}) and (\ref{INT}):

\begin{equation}
f(\theta;\Lambda, h) \equiv {e^{-3\pi i/4}\over
q_{0}^{\beta+1}}\sqrt{{2\pi \over
q_{0}}}\left[{\cosh^{2} q_{0}h  \over
3\sinh q_{0}h \cosh q_{0}h + q_{0}h}\right]
\sum_{n,m=0}^{\infty}
A_{m}^{(n)} J_{2m+n+\beta}(q_{0})\cos
n\theta .\label{Fh}
\end{equation}

The total scattering section is denoted $\Sigma$ and is
given by

\begin{equation}
\Sigma(q_{0},\Lambda, h) \equiv \int_{0}^{2\pi} d\theta
\vert f(\theta) \vert^{2} = {2 \pi^{2} \over
q_{0}^{2\beta+3}}\left[{
\cosh^{2} q_{0}h \over 3\sinh q_{0}h
\cosh q_{0}h +q_{0}h}\right]^{2}\sum_{n=0}^{\infty}\!\!'
\,\rule[-3mm]{.25mm}{8mm}\sum_{m=0}^{\infty}
A_{m}^{(n)}J_{2m+n+\beta}(q_{0})\,
\rule[-3mm]{.25mm}{8mm}\,^{2},\label{Sigmah}
\end{equation}

\noindent where $\sum'$ denotes multiplying by $2$
for the $n=0$ term.  Equations (\ref{Fh}) and
(\ref{Sigmah}) are the principle results of this section.

\subsection{Numerical Evaluation of $A_{m}^{(n)}$}

The linear set of equations (\ref{LIN1}) was solved by
numerically integrating the functions in (\ref{L2}) to
obtain complex matrix coefficients and then, for each
$n$, an $M\times M$ linear system was solved to
determine $A_{m}^{(n)}.$  All numerical integrations
demanded relative error $10^{-8}$. The dependence of
the $A_{m}^{(n)}$'s on the matrix size was found to be
negligible ($<10^{-6}$) provided $M\geq 14$.  $\beta$
was chosen to guarantee convergence of the integrals and
we chose $\beta = 1$ for our calculations as  this
appeared to give the fastest convergence for large values
of $m,l$. A numerical integration technique for infinite
integrals involving products of Bessel functions,
developed recently by Lucas,\cite{LUCAS} was used for the
integrations of the kernels in (\ref{L2}).

\subsection{Wave scattering on a heterogeneous free
membrane}

Since we are not aware of an analogy to capillary wave
scattering from a boundary condition inhomogeneity as
studied above, it is useful to consider a related scattering
problem with a similar geometry. In particular, we treat
wave propagation on a stretched elastic membrane and
calculate the scattering due to a circular region with
tension $T_{1}$ which is imbedded in a two-dimensional
membrane with tension $T_{0}$.  Vertical displacements
of this membrane obey the usual wave equation.  In
Section IV the scattering from this model is compared
with that derived in the previous section.

The configuration described in this section can in
principle be realized experimentally by embedding a thin
flexible loop of thread in a soap film. Surfactant can then
be added to the interior of the loop to decrease the
surface tension within the circle.  The thread now
supplies the line tension $\gamma$, which acts in the
tangent plane of the interface and satisfies

\begin{equation}
{\gamma \over a}= (T_{0}-T_{1}).
\end{equation}

\noindent The effects of an underlying liquid substrate
are not present in this example and the only dynamical
variable of the membrane is the height fluctuation
$\eta(r, \theta, t)$ which obeys a wave equation
$\partial_{t}^{2}\eta = c^{2}\nabla_{\!\perp}^{2}\eta$
with a dispersion relation

\begin{equation}
q_{j}^{2} = {\omega^{2} a^{2}\over c_{j}^{2}} \quad
\quad \quad
(j=0,1),
\end{equation}

\noindent where we have used the same rescaled
dimensionless quantities as in the previous section ({e.  g.,
$q=ka$). The wave speeds for the different regions are
$c_{j} = \sqrt{T_{j} / \rho_{m}}$ where $\rho_{m}$
denotes the areal mass density of the thin film.

We now wish to consider scattering of an incident plane
wave, $e^{iq_{0}r\cos \theta - i\omega t}$, by this
circular region of different tension.  The membrane
displacement, $\eta e^{-i\omega t}$, outside the circular
region takes the form

\begin{equation}
\begin{array}{ll}
\eta_{out}(r, \theta) = e^{iq_{0}r\cos \theta }
+\eta_{s}(r,\theta)  & \quad  \quad\quad \quad
(r>1) \label{OUT} \end{array}
\end{equation}

\noindent where the outgoing scattered height can be
expanded as

\begin{equation}
\eta_{s}(r,\theta) = \sum_{n=0}^{\infty} A_{n} H_{n}^{(1)}(q_{0}r)
\cos n\theta. \quad\quad \quad \quad (r>1)
\end{equation}

\noindent Similarly, the membrane displacement
inside the circle can be expanded as

\begin{equation}
\eta_{in}(r,\theta)=\sum_{n=0}^{\infty} B_{n} J_{n}(q_{1}r) \cos
n\theta. \quad\quad \quad \quad (0 \leq r <1)\label{IN}
\end{equation}

\noindent At the domain boundary $(r=1)$, continuity of
the sheet requires

\begin{equation}
\eta_{out}(1,\theta) = \eta_{in}(1,\theta),\label{C1}
\end{equation}

\noindent which also implies $\partial_{\theta}\eta$ is
continuous across $r=1$. The second boundary condition
requires careful consideration of the directionality of the
line tension.  Since both the interfacial tensions $T_{j}$
and the line tension $\gamma$ act in the surface of the
membrane, a balance of forces in the $z$ direction
requires continuity of slopes

\begin{equation}
\partial_{r}\eta_{out}(1,\theta) =
\partial_{r}\eta_{in}(1,\theta).\label{C2}
\end{equation}

\noindent Substituting the expressions for the height
displacements inside and outside the loop into the two
conditions (\ref{C1}) and (\ref{C2}) yields

\begin{equation}
A_{n} =
(2-\delta_{n,0})i^{n}{q_{0}J_{n}(q_{1})
[J_{n+1}(q_{0})-J_{n-1}(q_{0})]-
q_{1}J_{n}(q_{0})[J_{n+1}(q_{1})-J_{n-1}(q_{1})] \over
q_{1}H_{n}^{(1)}(q_{0})[J_{n+1}(q_{1})-J_{n-1}(q_{1})]
-q_{0}J_{n}(q_{1})[H_{n+1}^{(1)}(q_{0})-H_{n-1}^{(1)}
(q_{0})]},  \label{An0}
\end{equation}

\noindent for $n\geq 0$. We note that the tension ratio
enters this result since $q_1 / q_0 = \sqrt{T_0 / T_1}$. At
large distances, $r\rightarrow \infty$,  the membrane
displacement (\ref{OUT}) is

\begin{equation}
\eta_{out}(r,\theta) \simeq e^{iq_{0}r\cos \theta} +
f(\theta){e^{iq_{0}r}\over \sqrt{r}}
\end{equation}

\noindent with

\begin{equation}
f(\theta; T_0 / T_1) = \sqrt{{2 \over \pi q_{0}}} e^{-i\pi/4}
\sum_{n=0}^{\infty}(-i)^{n}A_{n}\cos n\theta.
\label{FT}
\end{equation}

The scattering cross section for this process is given by

\begin{equation}
\Sigma(q_{0}, T_0 / T_1)= {4\over q_{0}}
\vert A_{0}\vert^{2} +{2 \over q_{0}}
\sum_{n=1}^{\infty}\vert A_{n}\vert^{2},
\label{SIGMA}
\end{equation}

\noindent which will be compared in Section IV with the
corresponding results obtained in IIC.

\section{Scattering With Bending Rigidity Discontinuities}

\subsection{Waves on a fluid coupled membrane with
bending rigidity}

In this section, we consider the case of a thin elastic plate
overlying a fluid whose motion is assumed irrotational
and satisfies (\ref{LAPLACE}). Imbedded in the plate is a
circular region with different elastic constants. This
circular domain thus acts to scatter an incoming
transverse (bending) vibration.  Physical manifestations of
this problem are membranes with phase separated
domains, with imbedded impurities or defects, or other
films with a patch of different flexural rigidity
\cite{PETROV}  (see Figure 1b).

We assume that the only normal restoring force at the
interface is an elastic resistance to bending which
modifies the boundary conditions (\ref{NO1})  and
(\ref{NI1}). Balancing the pressure,
$p= -\rho\partial_{t}\phi$, the bending forces, and the
inertia of the membrane with mass density $\rho_{p}$
(\ref{BCN}), the equation for the
plate displacement $\zeta$ is\cite{LANDAU}

\begin{equation}
[D\nabla_{\perp}^{4}+
\rho_{p}\partial_{t}^{2}]\zeta(R,\theta)=p.\label{BCPLATE}
\end{equation}

\noindent Here $D$ is the bending rigidity, which for an
isotropic harmonic solid is typically defined
as\cite{LANDAU}

\begin{equation}
D_{j} = {E_{j} d^{3} \over 12(1-\mu_{j}^{2})} ,\label{D}
\end{equation}

\noindent where $d$ is the plate thickness, and $E_{j}$
and $\mu_{j}$ are Young's moduli and Poisson's ratios,
respectively. The elastic parameters will be assumed to
have different, though uniform,  values inside and outside
the circular domain. In microscopic systems such as
monolayers and bilayers where the ``plate'' thickness has a
molecular scale, the bending rigidity is typically denoted
$\kappa$, but the physical interpretation of $\kappa $ in
terms of material parameters as given  by (\ref{D}) is not
valid.

Taking a time derivative and using the kinematic
constraint (\ref{KIN}), we obtain the dynamic boundary
condition

\begin{equation}
[D\nabla_{\perp}^{4}\partial_{Z}+(\rho+
\rho_{p}\partial_{Z})\partial_{t}^{2}]\phi(R,\theta,Z)
\vert_{Z=0}=0.\label{BCPLATE1}
\end{equation}

The solution of this problem is similar to that of the
surface tension discontinuity. However, rather than
considering the general case of arbitrary values of fluid
density and membrane density, we simplify the problem
and consider two limiting, important cases. The first case
is when the normal forces from fluid pressure dominate
the plate's own inertial response. The second situation,
discussed in subsection B, neglects the effects of the
underlying fluid.  We note an important difference
between these two cases.  In the formulation for the fluid
coupled, massless plate, no further conditions on the
behavior of the fields at $r=1, z=0$ can be imposed. Here,
fluid velocities, surface displacements, and their
derivatives are continuous.  However, the physical model
considered may require extra conditions at $r=1,z=0$
which cannot be satisfied by the velocity potential of the
irrotational fluid.  These extra boundary conditions are
part of plate theory and are incorporated in the second
case, the scattering of free plate vibrations.

When the plate mass is neglected, the dispersion relation
derived from (\ref{BCPLATE}) is

\begin{equation}
q_{j}^{5}\tanh q_{j}h = {\rho \omega^{2}a^{5} \over D_{j}}.
\label{DISP1}
\end{equation}

\noindent Equation (\ref{DISP1}) is the massless plate
limit, $\rho_{p}/\lambda \ll \rho$, which, upon solving
for $\lambda = 2\pi/k$ using (\ref{DISP1}) for
$h=\infty$, is consistent with

\begin{equation}
\rho_{p} \ll \left({\rho^{4}D \over \omega^{2}
a}\right)^{1/5}.
\end{equation}

The detailed scattering calculation has the identical form
to that of the surface tension discontinuity problem
except that the functions $F(q)$ and $G(q)$ are changed
and the coefficients (\ref{L}) are modified. It is
straightforward to obtain

\begin{eqnarray}
L_{ml}^{(n)}(q_{0},h,\Lambda_d;x) &&=
i\pi(1-\Lambda_{d})\left[{\sinh q_{0}h
\cosh q_{0}h \over 5\sinh q_{0}h \cosh q_{0}h
+q_{0}h}\right] J_{2m+n+\beta}(q_{0})
J_{2l+n+\beta}(q_{0})  \nonumber \\&&
+\int_{0}^{1}dx
\left[I_{ml}^{(n)}(q_{0},h,\Lambda_d;x)+I_{ml}^{(n)}(q_{0},h,
\Lambda_d;2-x)\right]
+\int_{2}^{\infty}dx I_{ml}^{(n)}
(q_{0},h,\Lambda_d;x)\label{Ld}
\end{eqnarray}

\noindent with

\begin{equation} I_{ml}^{(n)}(q_0,h,\Lambda;x) \equiv
x^{1-2\beta}{x^{5}\tanh q_{0}hx -\Lambda_{d}\tanh
q_{0}h \over x^{5} \tanh q_{0}hx - \tanh q_{0}h}
\end{equation}

\noindent and  $\Lambda_{d} \equiv D_{0}/D_{1}$
denotes the ratio of outer to inner bending rigidities.  The
scattering amplitude $f$ and total scattering cross
section $\Sigma$ then follow from  (\ref{Fh}) and
(\ref{Sigmah}) by solving for the necessary coefficients
using (\ref{LIN1}), (with $\Lambda_{d}$ replacing
$\Lambda$), and (\ref{Ld}).

\subsection{Bending waves of a free plate}

Here it is instructive to study scattering in the uncoupled
plate limit. Provided $a \rho_{p}\omega^{2/5} \gg
\rho^{4/5}D^{1/5}$,  the pressure from the underlying
fluid can be neglected in the force balance
(\ref{BCPLATE}) and it is no longer necessary to consider
the underlying fluid (and so the solution for the velocity
potential).  Setting $p=0$ in (\ref{BCPLATE}) yields a wave
equation for $\zeta$ with the dispersion relation

\begin{equation}
\begin{array}{ll}
\displaystyle
q_{j}^{4}= \frac{\rho_{p}d\omega^{2} a^4}{D_{j}} &
\quad \quad j=0,1
\end{array}
\end{equation}

\noindent for waves outside ($j=0$) and inside ($j=1$)
the domain.  We treat the dynamics in a manner typically
used in plate theory, where shear stresses and bending
moments within the plate are balanced.\cite{PLATE}

As in Section IIE, the solution for $\zeta$ outside and
inside the circular domain can be expressed as

\begin{equation}
\begin{array}{ll}
\displaystyle
\zeta_{out} = e^{iq_{0}r\cos \theta -i \omega t} +
\zeta_{s}(r,\theta)e^{-i\omega t} & \quad \quad (r >1)
\label{fout} \end{array}
\end{equation}

\noindent where

\begin{equation}
\begin{array}{ll}
\displaystyle
\zeta_{s}(r,\theta) =
\sum_{n=0}^{\infty}\left[A_{n}H_{n}^{(1)}(q_{0}r)+B_{n}
K_{n}(q_{0}r)\right] \cos n\theta\label{fscatt}
& \quad (r>1),
\end{array}
\end{equation}

\noindent and

\begin{equation}
\begin{array}{ll}
\displaystyle \zeta_{in}(r,\theta) =
\sum_{n=0}^{\infty}\left[C_{n}J_{n}(q_{1}r)+E_{n}I_{n}
(q_{1}r)\right] \cos n\theta & \quad\quad (0\leq r <1) .
 \label{fin}
\end{array}
\end{equation}

To determine the coefficients $A_{n}, B_{n}, C_{n},$ and
$E_{n}$, we apply four constraints at the boundary $r=1$:
continuity of the plate displacements, the pure bending
approximation, balance of vertical shear forces, and
balance of rotational moments.  These quantities can be
expressed in the following terms: $\zeta_{in}^{-} =
\zeta_{out}^{+}, \beta_{r}^{-} = \beta_{r}^{+},
M_{rr}^{-}=M_{rr}^{+}$, and $V_{rz}^{-}=V_{rz}^{+},$
where $\beta_{r}, M_{rr}$, and $V_{rz}$ are the bending
angle, the bending moment, and the Kirchhoff shear
stress resultant respectively, and $\pm$ denote the
functions evaluated as $r\rightarrow 1$ from outside and
inside, respectively.\cite{LANDAU,PLATE} Expressing these
four conditions in terms of the out-of-plane
displacements, we conclude that

\begin{equation}
\zeta, \quad \, {\partial \zeta \over \partial r},\quad\,
D(r)\left[\nabla^{2}\zeta+(\mu(r)-1)\left({1 \over
r}{\partial \zeta\over \partial r}+{1 \over
r^{2}}{\partial^{2}\zeta \over \partial
\theta^{2}}\right)\right], \label{BCP0}
\end{equation}

\noindent and

\begin{equation}
D(r)\left[ \nabla^{2}{\partial \zeta \over \partial r}
+ {\mu(r)-1\over r^{2}}
\left({1 \over r}{\partial^{2}\zeta \over \partial
\theta^{2}} - {\partial^{3}\zeta \over \partial \theta^{2}
\partial r}\right)\right]\label{BCP1}
\end{equation}

\noindent are all continuous across the boundary of the
circular inclusion. The latter two boundary conditions at
the domain boundary are not enforced for the fluid
coupled problem, as discussed in the next section.
Substituting the expansions (\ref{fout}-\ref{fin}) into
these four boundary conditions yields a linear set of
equations which determine the unknown coefficients.  In
particular, we find

\begin{equation}
A_{n} =   {det \,{\bf N} \over det \,{\bf M}}, \label{An}
\end{equation}

\noindent where the matrices ${\bf M}$ and ${\bf N}$ are
given in Appendix B.  Expressions for the scattering
amplitude and total scattering cross section, which are
functions of the $A_{n}$,  are identical to those of the
tension discontinuity problem and are given by equations
(\ref{FT}) and (\ref{SIGMA}).

\section{Discussion}

\subsection{Results}

We first examine capillary wave scattering on a
heterogeneous surface using equations (\ref{Fh}) and
(\ref{Sigmah}), and compare with scattering in a free
membrane, equation (\ref{SIGMA}).  Polar plots of the
scattering amplitude, $\vert f(\theta, h=\infty)\vert $, for
various incoming wavevectors $q_{0}$ and $\Lambda
=2$ are shown in Figure 3.  In the long wavelength limit,
$q_{0} \ll 1$, the domain is nearly a point scatterer, and
the scattering is nearly isotropic as expected; even for
$q_{0}\approx 1$, there is almost no variation with
$\theta$.  Distinct lobe structures develop when the
incoming wavelength is about three times the radius of
the scattering domain.  Figures 4a-d illustrate,
respectively, the effect of subphase depth on scattering
for wavevectors $q_{0}=1.0,2.0,3.0,$ and $4.0$, again for
$\Lambda =2$.  In each case, depths $h=0.1, 0.5$, and
$h=\infty$ are considered. For these wavevectors,
decreasing $h$ decreases the scattering amplitude for all
$\theta$, but has almost no effect on the angular
variations.

The  total scattering cross sections as a function of the
reduced incident wavevector  for $\Lambda=2.0$ and $
10.0$ are shown for various depths in Figures 5a,b
respectively. The depth dependence of $\Sigma$ is
shown in Figure 6.  Figures 5a,b show oscillations in
$\Sigma$ as $q_{0}$ is varied.  Unlike examples of sound
wave scattering from an infinitely rigid
obstacle,\cite{MORSE} where a monotonic variation of
$\Sigma$ with $q_{0}$ is expected,  the calculations
presented here exhibit a different and more complex
dynamical behavior. \cite{MH1} The curves for $\Sigma$
have oscillations and maxima analogous to Mie
resonances in the scattering of radiation from a dielectric
sphere.\cite{MIE} In Figure 6, a weak variation of
$\Sigma$ with $h$ is observed which varies most
significantly for intermediate $q_0$.  We note that for
shallow underlying fluids viscous dissipation is
important when $\nu \gg a^{2}\omega h^{2}$, and is
discussed further below.  We have also compared our
results with the asymptotic expression in the small
wavenumber limit ($q_0\ll 1$), $\Sigma = \frac{\pi^2}{9}
(\Lambda-1)^2 q^{3}$, developed by Chou and
Nelson.\cite{TC2} The agreement is good as displayed in
Figure 5c.

We compare the above results with those of scattering in
a free membrane (the effects of the underlying fluid are
neglected, section IIID),  shown as the darker dashed
curves in Figures 5a,b. Although these two problems are
geometrically similar, they have different dispersion
relations, so in order to provide a meaningful comparison,
we study two cases with the ratio of wavevectors
$q_{1}/q_{0}=2^{1/3}$ and $10^{1/3}$ ($\Lambda=
\sigma_{0}/\sigma_{1}=2,10$ but $T_{0}/T_{1}= 2^{2/3},
10^{2/3}$).  A plot of (\ref{SIGMA}) for $T_{0}/T_{1}=2$
would correspond to a scale change (compression) along
the $q_{0}$ axis in Figure 5.  The dependence of
$\Sigma$ on $q_{0}$ for scattering in the uncoupled
membrane versus the capillary wave problem is
qualitatively similar at the same ratio of inner to outer
wavevectors. In fact, the scattering cross section for
$\Lambda=2.0, h=\infty$ in Figure 5a is nearly
indistinguishable  from  that of the free membrane. The
corresponding amplitudes $\vert f(\theta,h) \vert$ for the
uncoupled membrane,  equation (\ref{FT}), are nearly
identical to the polar plots shown in Figure 4
corresponding to $\Lambda=2.0$.  For $\Lambda=10.0$,
Figure 5b, the free membrane model is now clearly
distinguishable from the capillary wave scattering case,
even after the wavevectors are rescaled to account for the
different dispersion relations between a free membrane
and a capillary wave.

To illustrate the effects of varying the ratios of the surface
parameters, we have plotted in Figure 7 $\Sigma(q_{0},
\Lambda, h=\infty)$ as a function of $\Lambda^{-1}$ for
$q_{0}=1.0, 2.0,$ and $3.0.$ The limit
$\Lambda^{-1}=\sigma_{1}/\sigma_{0} \rightarrow
\infty$ is associated with an infinitely ``hard'' domain at
the interface which resists increases in surface area,
though we note that such monolayer systems are
unstable and can exist only in the presence of an
externally applied force. Increasing $\Lambda$ from
unity corresponds to a more flexible domain, which
actively supports internal fluctuations, and leads to a
more complicated scattering and far-field disturbance.

We now present results  for scattering in interfaces
characterized by bending rigidity rather than surface
tension.  Scattering amplitudes for the fluid-coupled
membrane for $q_{1}/q_{0}=2^{1/3}$ (corresponding to
$D_{0}/D_{1}=2^{5/3}$) are shown in Figures 8a-d for
$q_{0}=1.0, 2.0,  3.0,$ and $4.0$ and different fluid
depths. Total scattering cross sections as a function of
$q_{0}$ are plotted in Figure 9.  For comparison, the
scattering amplitude and cross sections for bending
waves of a free plate (no underlying fluid)  are depicted
by the dotted curves in both Figures 8 and 9.  To maintain
scale similarity in the internal to external wavelength
ratio, $q_{1}/q_{0} = 2^{1/3}$,  we choose $D_{0}/D_{1} =
2^{4/3}$.  In addition, due to the elastic boundary
conditions  at $r=1$ in the plate model, we need to
specify the contrast in both bending rigidity $D$ and
Poisson's ratio $\mu$. The dotted curve in Figure 9
was calculated with $\mu_{0}=0.25,
\mu_{1}=-0.25.$\cite{NELSON} From other numerical
simulations, we have found that the cross sections
$\Sigma$ are rather insensitive to the discontinuity in
Poisson's ratios except for small variations in the fine
structure.

The discontinuities present in the higher derivative
boundary conditions (\ref{BCP0}) and (\ref{BCP1}) for
the free elastic plate make this problem
mathematically different from that of the massless
bending interface coupled to an underlying fluid,
where the velocity potential is assumed harmonic.  For
the fluid coupled irrotational flow problem, only two
boundary conditions at $r=1$ can be satisfied.
Solution of the full boundary value problem with four
boundary conditions applied at the domain boundary,
(\ref{BCP0}) and (\ref{BCP1}), requires consideration of
viscous effects in the underlying fluid.  The
discontinuities at $r=1$ suggested by (\ref{BCP0}) and
(\ref{BCP1}) are smoothed out in the thin boundary
layer with thickness $\sqrt {\nu/\omega}$ adjacent to
the upper surface.  It appears that the differences in
Figures 8 and 9 for wave scattering along the
heterogeneous plate as compared with the
comparable capillary wave scattering problem are the
discontinuous boundary conditions at $r=1$.  In the
surface tension problems solved in Section II, the
membrane displacement and its first derivative are
both continuous and so both (\ref{C1}) and (\ref{C2})
are satisfied.

The oscillations in $\Sigma(q_{0})$ for  surface
tension scattering, bending rigidity scattering, and
scattering in an uncoupled membrane or plate,
(Figures 5 and 9), are similar to those of the cross
section of electromagnetic radiation scattering from a
dielectric sphere. \cite{MIE} In that case the radiation
field penetrates a sphere with finite dielectric
constant; internal reflections constructively or
destructively interfere depending on the incoming
wavelength. For the problem studied here, capillary
wave scattering from an interfacial inhomogeneity,
internal waves are supported, and interferences lead
to the oscillations in $\Sigma(q_{0})$.

\subsection{Experimental Consequences}

Our analysis has been confined to wave scattering from
an isolated domain. However, under most experimental
conditions the morphology of the monolayer film cannot
be directly controlled. For example, when amphiphilic
molecules are deposited at the liquid-air interface they
form a film which is in a phase separated state consisting
of a collection of scatterers of different sizes.  The density
and sizes of these domains depend on surfactant
concentration.\cite{REV,REV2} Interfacial surfactant
morphology thus affects capillary wave propagation with
scattering as a source of enhanced wave
attentuation.\cite{NOSKOV,WANG}

Experiments on capillary wave damping
\cite{WANG,LUC,EXP1,EXP2,EXP3} show at least one
maximum in the attenuation coefficient as a function
of total surface concentration of surfactant.  Wave
amplitudes, measured by $|\eta|^2$, can be related to
the scattering in the dilute scatterer limit (single
scattering only) by Beer's law\cite{ISHIMARU}

\begin{equation}
\vert \eta(x) \vert^{2} \propto e^{-c \Sigma(q_{0})x}
\label{BEER}
\end{equation}

\noindent where $c$ is the area concentration of
domains of radius $a$. Since the calculated cross
section $\Sigma(q_{0})$ contains peaks (Fig. 5a,b and
Fig. 9) (\ref{BEER}) at least qualitatively accounts for
the maxima in the damping coefficients which are
experimentally measured.\cite{LUC,EXP1,EXP2,EXP3}
Equation (\ref{BEER}) suggests that peaks may
observed as $q_{0}$ is varied.  Indeed, the variation in
damping as domain size and incoming wavevector are
tuned has been qualitatively observed in the
experiment of Wang {\it et.  al.}\cite{WANG} who
measured a sharp increase in damping as the
incoming wavelength was decreased from $\lambda
\gg a$ to $\lambda \approx a$.

We conclude the discussion by examining the region
of validity of the analytical results derived in Sections
II and III.  Throughout the analysis, effects of viscosity
have been neglected. This approximation is valid
provided the domain is larger than the viscous length,
$\sqrt{\nu /\omega},$ {\it i.e.,} the Stokes layer
adjacent to the fluctuating boundary. Thus we require

\begin{equation}
a \gg \sqrt{{\nu \over \omega}} \label{CONDa}
\end{equation}

\begin{equation}
H \gg \sqrt{{\nu \over \omega}}.
\label{CONDS}
\end{equation}

\noindent When the dispersion relation (\ref{DISP0}) for
infinite depth is used to eliminate $\omega$ from
(\ref{CONDS}), and a condition that the waves are not too
heavily damped by viscosity is imposed, $k_j^{2} \ll
\omega /\nu$, we obtain validity criteria for $q_j=k_j a$,

\begin{equation}
\displaystyle \left({a^2 \omega \over \nu}\right)^{1/2}
\gg q_j \gg \left({\nu^{2}\rho
\over a \sigma_{j}}\right)^{1/3}. \label{CONDQ}
\end{equation}

\noindent  For an air-water interface, $\sigma \approx $
70 dynes/cm, $\nu \approx $ 0.01 cm$^{2}$/sec, and
$\rho \approx $1 gm/cm$^{3}$. The first relation in
(\ref{CONDQ}) provides an upper bound of $k_{j} = q_{j}/a
\ll \sigma /\rho \nu^{2} \approx 7 \times 10^{5}$
cm$^{-1}$ and a lower bound of $k_{j}= q_{j}/a \gg 113$
cm$^{-1}$, 5.2 cm$^{-1}$, and 0.24 cm$^{-1}$ for
domain radii of $a\approx 10\mu$m, $100\mu$m, and
1mm respectively.

For a rigid membrane described by the dispersion
relation (\ref{DISP1}), the conditions $k_j^{2} \ll
\omega/\nu$ and (\ref{CONDa}) lead to

\begin{equation}
k_{j}a=q_{j}\gg max\left[ \left({a\nu^{2}\rho \over
D_{j}}\right)^{1/5}, {a\nu^{2}\rho \over D_{j}},
\,\left( {g\rho \over D_{j}}\right)^{1/4} \right]
\label{DCON}
\end{equation}

\noindent where the last term is a condition due to the
neglect of gravity waves. For typical estimates of the
bending rigidity in cell membranes,\cite{PAL} $D\sim
10^{-12}$ergs, the above condition (\ref{DCON})
requires exceedingly small wavelengths, $\lambda \ll
1$nm, to be satisfied.  As an estimate of the range of
validity, consider a system with $D=0.1$ ergs. For
domain radii of $10\mu$m, $100\mu$m, and
$1$mm, (\ref{DCON}) requires $k_{j} \gg 63 $
cm$^{-1}$, 10 cm$^{-1}$, and 9.95 cm$^{-1}$,
respectively.

\section{Summary}

In this paper, we have formulated an analytical method
for studying surface wave scattering due to a circular
inhomogeneity in the boundary conditions.  The results of
the calculations imply that scattering of capillary waves
from a circular domain where there is a discontinuity in
surface properties (surface tension or bending rigidity) is
not very sensitive to the depth of the underlying liquid
(see Fig. 7).  The scattering patterns $\vert
f(\theta;h)\vert$ have lobed structures indicative of
angular variations in the far-field disturbance. In
addition, an examination of the above results (Figures 5a
and 9) show that differences between scattering from a
surface tension contrast and a bending rigidity contrast
are small if the wavevector ratio, $q_{0}/q_{1}$, of the
two problems is maintained constant.  For a thin elastic
plate, the scattering from a domain of different rigidity
has some qualitative differences when compared with the
scattering from the fluid coupled system (Figure 9); this
difference is due to the additional  boundary conditions
applied at $R = a$ for the free plate problem whereas the
neglect of viscosity requires the elimination of two
boundary conditions in the irrotational flow problem.

Of course, in real experiments of phase separated
surfactant films, the domains are polydisperse in size,
often interact, and may not be circular. To incorporate
such effects, a statistical average over domain sizes and
correlations in the scattered potential must be included.
However, we expect the qualitative behavior of the simple
models described here to be representative of the
physical features of these systems. Furthermore, phase
separating monolayers are rarely in equilibrium, the
typical sizes and number density change after the
ambient conditions are changed, and the time evolution
of the film obeys complicated coarsening and topological
relaxation mechanisms. It is not unreasonable that
experimentally, long times need to elapse before
reproducible damping results are attained.  In fact, when
scattering is important, capillary wave damping might be
used as a diagnostic for probing the statistical averages
and effective influences of phase separating
morphologies.

The calculations presented for scattering with irrotational
flows can be extended to include systems of membranes
completely immersed in a fluid.  Common examples are
cell membranes, emulsions, or bilayers. As long as the
geometry is symmetric top and bottom, both side
contribute equally to the dynamical pressure; thus, the
normal stress boundary condition becomes
$\sigma(\vec{r})\nabla_{\!\perp}^{2}\eta =
2\rho\partial_{t}\phi.$ All the remaining steps follow
with the substitution $\rho\rightarrow 2\rho$. Results of
scattering for such an immersed membrane are simply
those for scattering at an air-liquid interface with
$q_{0},q_{1}$ rescaled by the modified ($\rho
\rightarrow 2\rho$) dispersion relation (\ref{DISP}).

We have only considered a circular domain which affects
the normal stress balance at the interface. However, the
mathematical methods utilized in this paper can be
extended to study tangential stress variations due to
viscoelastic properties intrinsic to the film.\cite{TC2}
Furthermore, a similar approach may be applicable for
studying the effects of nonuniform surface stresses on
the propagation of surface acoustic waves.

\acknowledgements

T. Chou acknowledges Dionisios Margetis, Vassilios
Houdzoumis, and David R.  Nelson for their helpful
discussions. H. A. Stone gratefully acknowledges a grant
from NSF-PYI Award CT5-8957043.

\appendix
\section{viscous effects}

In this appendix we reconsider capillary wave scattering
from a circular surface tension discontinuity in the
presence of substrate fluid viscosity. We study solutions
of the continuity equation $\nabla\!\cdot\!\vec{v}=0$ and
the time-dependent Stokes equation, $\partial_{t}\vec{v}
= -\nabla (p/\rho) + \nu \nabla^{2}\vec{v}$ for small
amplitude capillary waves in an infinite depth fluid. Here,
all variables $r, z,$ and $t$ are dimensional. As before,
the system is driven at a fixed frequency $\omega$ and
all quantities are  assumed to have  an $e^{-i\omega t}$
time dependence. The limit of small viscous effects
corresponds to the analysis presented in Section II.

When considering viscosity of the underlying liquid,
tangential stresses at the interface must also be balanced.
Accounting for the line tension $\gamma$ at the circular
domain boundary and changes in surface tension
$\sigma$, and neglecting any surface viscosity or
elasticity of the surface film, the tangential stress balance
has the form

\begin{equation}
-{\gamma \over
a}\delta(r-a)+\vec{\nabla}_{\!\perp}\sigma(r) +\rho\nu
(\partial_{z} \vec{v}_{\perp} + \vec{\nabla}_{\!\perp}
v_{z})\, \rule[-2mm]{.2mm}{6mm}\,_{\stackrel{ }{z=0}}= 0.
\label{T1}
\end{equation}

For a step change of surface tension treated here
(equation \ref{s0s1}), the discontinuity of $\nabla_{\perp}
\sigma(\vec{r})$ at $r=a$ is balanced by the line tension
term, ${\gamma \over a}\delta(r-a)$ and the surface
tangential stress condition obeyed everywhere on the
interface is

\begin{equation}
(\partial_{z}
\vec{v}_{\perp} + \vec{\nabla}_{\!\perp} v_{z})\,
\rule[-2mm]{.2mm}{6mm}\,_{\stackrel{ }{z=0}}
=0 .\label{TAN}
\end{equation}

\noindent The normal stress balance including viscous
effects is

\begin{equation}
\sigma(\vec{r})\nabla_{\!\perp}^{2}\eta
= - p +2\rho\nu\partial_{z}v_{z} \,\rule[-2mm]
{.2mm}{6mm}\,_{\stackrel{ }{z=0}} \label{A3}
\end{equation}

\noindent Taking a time derivative of (\ref{A3}) and using
(\ref{KIN}), we have

\begin{equation}
\sigma(\vec{r})\nabla_{\perp}^{2}v_{z}
+2i\rho\nu\omega\partial_{z}v_{z}= i\omega
p\,\rule[-2mm]
{.2mm}{6mm}\,_{\stackrel{ }{z=0}} \label{A4}
\end{equation}

\noindent To solve the unsteady Stokes equations, it is
convenient to decompose the velocity fields as \cite{TC2}

\begin{equation}
\vec{v}=\vec{\nabla} \phi(r) + \left( \begin{array}{c}
\vec{u}_{\!\perp} \\
u_{z}\end{array} \right), \label{DECOMP}
\end{equation}

\noindent where $\nabla^{2}\phi=0$, and
$\vec{u}_{\perp}$ and $u_{z}$ satisfy the diffusion
equation (recall all fields are proportional to
$e^{-i\omega t}$)

\begin{equation}
-i\omega \vec{u}_{\!\perp} = \nu
\nabla^{2}\vec{u}_{\!\perp} \quad \quad
-i\omega u_{z} = \nu
\nabla^{2}u_{z}.
\label{DIFF}
\end{equation}

\noindent The pressure is $p=i\omega\rho \phi$.

We treat the incident $(i)$ and scattered $(s)$ waves
separately by defining the incident fields, $\phi^{(i)}(x,z)$
and $(\vec{u}_{\!\perp}^{(i)}, u_{z}^{(i)})$, and scattered
fields, $\psi(r,\theta,z)$ and $(\vec{u}_{\!\perp}^{(s)},
u_{z}^{(s)})$, such that

\begin{equation}
\begin{array}{l}
\phi(r,\theta,z)  = \phi^{(i)}(x,z)  + \psi(r,\theta,z) \\
\vec{u}_{\!\perp}  = \vec{u}_{\!\perp}^{(i)}(x,z)
+ \vec{u}_{\!\perp}^{(s)}(r,\theta,z) \\
u_{z} = u_{z}^{(i)}(x,z) + u_{z}^{(s)}(r,\theta,z).
\end{array}
\end{equation}

\noindent The incident fields $\phi(x,z)$ and
$(\vec{u}_{\!\perp}^{(i)}, u_{z}^{(i)})$ determine the fluid
velocity of a damped wave on a uniform interface of
surface tension $\sigma_{0}$. These fields must satisfy
the normal stress boundary condition (\ref{A4}) with
$\sigma = \sigma_{0}$ and the two-dimensional
divergence of the tangential stress boundary condition
(\ref{A3}), respectively,

\begin{equation}
\begin{array}{l}
\sigma_{0}\nabla_{\perp}^{2}u_{z}^{(i)} +
\rho\omega^{2}\phi^{(i)}+2i\omega\nu
\partial_{z}u_{z}^{(i)}= 0 \\
2\partial_{z}\nabla_{\perp}^{2}\phi^{(i)} -
\partial_{z}^{2} u_{z}^{(i)}+\nabla_{\perp}^{2}
u_{z}^{(i)} = 0.\label{NT0}
\end{array}
\end{equation}

\noindent Consider an incident wave with a velocity
potential of unit amplitude on the $y$-axis, $x=z=0.$
Substitution of $\phi^{(i)}(x,z) = e^{iQ_{0}x+Q_{0}z}$ and
$u_{z}^{(i)}(x,z) = A e^{iQ_{0}x+l_{0}z}\!,$
where $l_{0}^{2} =
Q_{0}^{2}-i\omega/\nu$, into equation (\ref{NT0}) yields
two equations for the unknowns $A$ and
$Q_{0}(\omega)$ which are satisfied when

\begin{equation}
A= - {2\nu Q_{0}^{3} \over 2\nu Q_{0}^{2} -i\omega}
\end{equation}

\noindent and

\begin{equation}
\sigma_{0}Q_{0}^{3}-\rho\omega^{2}-
4\rho\nu^{2}Q_{0}^{2} l_{0}(Q_{0}+l_{0})
= 0.\label{POLY}
\end{equation}

The condition (\ref{POLY}) determines the dispersion
relation for capillary waves on a homogeneous surface
with tension $\sigma_{0}$. In the $\nu \rightarrow 0$
limit, the first correction to the pole at
$\sigma_{0}k_{0}^{3} = \omega^{2}a^{3}$
gives\cite{LEVICH}

\begin{equation}
Q_{0}  \approx k_{0} + i\epsilon \equiv k_{0} + i{4
\omega \nu \over 3\sigma_{0}}, \label{POLE}
\end{equation}

\noindent which shows that incident waves are
spatially damped by the factor
$e^{(-4\omega\nu/3\sigma_{0} )x}$.

We now consider scattering from a domain of radius
$r=a$ and surface tension $\sigma_{1}$.  The incident
wave is the same as in the above nonscattering case,
{\it i.e.}, a wave that has unit velocity potential
amplitude in the absence of the scatterer.  The normal
stress condition obeyed by the scattered fields is

\begin{equation}
\begin{array}{ll}
{\sigma_{0}\over \rho}\nabla_{\perp}^{2}u_{z}^{(s)}+
\omega^{2}\psi+2i\omega\nu
\partial_{z}u_{z}^{(s)} = 0 &
\quad \quad \quad\quad (r>1) \\
{\sigma_{1}\over \rho}\nabla_{\perp}^{2}u_{z}^{(s)}+
\omega^{2}\psi+2i\omega\nu
\partial_{z}u_{z}^{(s)}  & \,\, \\
\hspace{6mm} =
-{\sigma_{1}\over \rho} \nabla_{\perp}^{2}u_{z}^{(i)}-
\omega^{2}\phi^{(i)}-2i\omega\nu
\partial_{z}u_{z}^{(i)} & \quad \quad
\quad (0 \leq r<1). \label{BCN1}
\end{array}
\end{equation}

\noindent Since the tangential stress condition holds
for all $r$, we can use (\ref{TAN}) to eliminate the
$u_{z}^{(s)}$ terms from
(\ref{BCN1}) hence solving for $\psi$.

Expansions for the fields $\psi$ and $(\vec{u}_{\!\perp},
u_{z})$ can be found which obey the Laplace and
diffusion equations respectively.  For each angular
harmonic $\cos m\theta$, $\psi(r,\theta,z)$ can be
written as a Hankel transform of $\psi_{m}(k)e^{kz}$ as
in equation (\ref{DEF}); similarly $u_{z}^{(s)}$ can be
expanded in terms of $A_{m}(k)J_{m}(kr)e^{lz}$.
Furthermore, with $\vec{u}_{\!\perp} \equiv (u_{r},
u_{\theta})$, it is convenient to work with
$u_{r}^{(s)}\pm u_{\theta}^{(s)}$, which are expanded in
terms of $U^{\pm}_{m}(k)J_{m\pm 1}(kr)e^{lz}$;

\begin{equation}
\psi(r,\theta,z) = \sum_{m=0}^{\infty}\cos m\theta
\int_{0}^{\infty}\!dk \,\psi_{m}(k) J_{m}(kr)
e^{kz},\label{psiJ}
\end{equation}

\begin{equation}
u_{r}^{(s)}(r,\theta,z) = \sum_{m=0}^{\infty}\cos m\theta
\int_{0}^{\infty}\!dk\,\left[U^{+}_{m}(k) J_{m+1}(kr)
+U^{-}_{m}(k) J_{m-1}(kr)\right] e^{lz}
\label{ur}
\end{equation}

\begin{equation}
u_{\theta}^{(s)}(r,\theta,z) =
\sum_{m=0}^{\infty}\sin m\theta
\int_{0}^{\infty}\! dk \,\left[U^{+}_{m}(k) J_{m+1}(kr)
-U^{-}_{m}(k) J_{m-1}(kr)\right] e^{lz}
\label{ut}
\end{equation}

\noindent and

\begin{equation}
u_{z}^{(s)}(r,\theta,z)  =
\sum_{m=0}^{\infty}\cos m\theta \int_{0}^{\infty}\!dk
\,A_{m}(k) J_{m}(kr) e^{lz}\label{uz}
\end{equation}

\noindent where $l \equiv \sqrt{k^{2}-i\omega /\nu}$
and we must now determine the four functions
$U_{m}^{\pm}(k), A_{m}(k),$ and $\psi_{m}(k).$ Note the
above integrands have a branch cut starting at
$k=\sqrt{i\omega/\nu}$.

The remaining steps follow those described earlier
though the algebraic details are more cumbersome.
Substitution of (\ref{psiJ}) and (\ref{uz}) into equation
(\ref{TAN}),  multiplying the integrand by $\cos n\theta$
and integrating $\theta$ from $0$ to $2\pi$, we relate
the two fields

\begin{equation}
A_{n}(k) = -{2\nu k^{3}\over 2\nu k^{2}
-i\omega}\psi_{n}(k).
\end{equation}

Applying the normal stress balance on the interface in the
interior of the circular scatterer yields the dual integral
equations of the same form as (\ref{DIE2}) where now
we find the dimensionless functions

\begin{eqnarray}
F_{n}(q;\tilde{Z}) = {i^{-n}\over
a}\left(q_{0}^{3}-q^{3}+4\tilde{Z}^{2}
q^{2}m(q+m)\right)
\psi_{n}(q/a)\label{Fnu}
\end{eqnarray}

\noindent and

\begin{eqnarray}
G(q;\tilde{Z}) =
{q^{3}-q_{1}^{3}-4\tilde{Z}^{2}\Lambda
q^{2}m(q+m) \over  q^{3}-q_{0}^{3}-4
\tilde{Z}^{2}\Lambda
q^{2}m(q+m)},\label{Gnu}
\end{eqnarray}

\noindent corresponding to (\ref{Fq}) and (\ref{Gq})
respectively, where the strength of the viscous effects are
measured by the Ohnesorge number

\begin{equation}
\tilde{Z}^{2} = {\rho \nu^{2} \over \sigma_{0} a}.
\end{equation}

\noindent We have used the definition (\ref{DISP}) in the
$h\rightarrow \infty$ limit for $q_{j}$ and defined the
dimensionless quantity $m=la = \sqrt{q^{2} -
q_{0}^{3/2}/\tilde{Z}}$. Also, the equivalent of the right
side of the second equation in (\ref{DIE2}) is

\begin{equation}
\Delta_{n}(Q_{0};\nu) J_{n}(Q_{0}r)\equiv
(2-\delta_{n,0})(1-\Lambda)\left[ {Q_{0}^{3}a^{3} \over
1+2iQ_{0}^{2}a^{2}\tilde{Z}/q_{0}^{3/2}}\right].
\end{equation}

\noindent Solving the the dual integral equations yields
$F_{n}(q;\tilde{Z})$ from which the $\psi_{n}(q/a)$ are
determined. When $\tilde{Z}$ small,  the capillary wave
pole in $G(q;\tilde{Z})$ is $q\approx Q_{0}$  as given in
(\ref{POLE}).  This positive imaginary shift when $\nu$
approaches zero fixes the integration path below the pole
at $q \approx k_{0}$ in (\ref{L}). In the $\nu\rightarrow
0\, (\tilde{Z}\rightarrow 0)$ limit, this problem reduces to
that of scattering of surface waves on an irrotational fluid
of infinite depth.

The scattered velocity fields can be extracted by using
the decompositions of the scattered fields,
(\ref{psiJ}-\ref{uz}) to balance the $\hat{r}$ and
$\hat{\theta}$ components of the tangential stress
equation (\ref{TAN}).  The remaining expansion
coefficients $U^{\pm}_{n}(k)$ are solved in terms of the
$\psi_{n}(k)$:

\begin{equation}
U^{\pm}_{n}(k)= \pm{\nu k^{2}l \over 2\nu
k^{2}-i\omega} \psi_{n}(k).
\end{equation}

\noindent Adding the contributions
from $\psi(r,\theta,z)$, the
scattered velocities can be formally written as

\begin{equation}
u_{r}^{(s)}(r,\theta,z) = \sum_{m=0}^{\infty}\cos m\theta \,
\int_{0}^{\infty}\!dk\left[e^{kz}-{2\nu kl \over 2\nu
k^{2}-i\omega}e^{lz}\right]{d J_{m}(kr) \over
dr}\psi_{m}(k)\label{Ur}
\end{equation}

\begin{equation}
u_{\theta}^{(s)}(r,\theta,z) = \sum_{m=0}^{\infty}\,
\sin m\theta \int_{0}^{\infty}\!dk\left[-e^{kz}
+{2\nu kl \over 2\nu k^{2} -i\omega}e^{lz}\right]
{m\over r}J_{m}(kr)\psi_{m}(k)\label{Ut}
\end{equation}

\begin{equation}
u_{z}^{(s)}(r,\theta,z) = \sum_{m=0}^{\infty}\cos m\theta
\int_{0}^{\infty} k dk \left[e^{kz} - {2\nu k^{2} \over
2\nu k^{2}-i\omega}e^{lz}\right] J_{m}(kr)\psi_{m}(k).
\label{Uz}
\end{equation}

\noindent To obtain the total velocity field of the
scattering process, the incident velocities, $\vec{v}^{(i)}$,
(corresponding to $\phi^{(i)}$ and $(u_{\!\perp}^{(i)},
u_{z}^{(i)})$ in (\ref{DECOMP})) must be added

\begin{equation}
v_{x}^{(i)} = iQ_{0} e^{iQ_{0}x + Q_{0}z}-i{Q_{0}\over
l_{0}}{2\nu Q_{0}^{3}
\over 2\nu Q_{0}^{2}-i\omega}e^{iQ_{0}x + l_{0}z}
\end{equation}

\begin{equation}
v_{z}^{(i)} = Q_{0} e^{iQ_{0}x + Q_{0}z}-{2\nu Q_{0}^{3}
\over 2\nu Q_{0}^{2}-i\omega}e^{iQ_{0}x + l_{0}z}.
\end{equation}

\noindent where $Q_{0}$ is given by the roots of
(\ref{POLY}). \cite{LEVICH}

\section{Boundary Conditions for Bending Waves}

This Appendix tabulates intermediate results for the
scattering produced by a circular inclusion with bending
rigidity $D_{1}$ imbedded in a membrane with bending
rigidity $D_{0}$. The four boundary conditions
formulated in Section III B lead to four linear equations
for the coefficients, $X_{i} = (A_{n}, B_{n}, C_{n}, E_{n})$
which may be compactly written

\begin{equation}
M_{ij}X_{j} = B_{i}
\end{equation}

\noindent where

\begin{equation}
{\bf B} = i^{n}(2-\delta_{n,0})
\left[ \begin{array}{c}
J_{n}(q_{0}) \\ J_{n}'(q_{0}) \\
(q_{0}^{2}+(\mu_{0}-2)n^{2})
J_{n}(q_{0})-q_{0}(\mu_{0}-1) J_{n}'(q_{0}) \\
n^{2}(\mu_{0}-1)J_{n}(q_{0})+q_{0}(q_{0}^{2}-
\mu_{0}n^{2})J_{n}'(q_{0})
\end{array}\right]
\end{equation}

\noindent and ${\bf M}$ is the $4\times 4$ matrix given
by

\begin{equation}
M_{1i} = \left(-H_{n}^{(1)}(q_{0}), -K_{n}(q_{0}), \,
J_{n}(q_{1}), \,I_{n}(q_{1})\right),
\end{equation}

\begin{equation}
M_{2i} = \left( -H_{n}^{(1)} {}'(q_{0}), -K_{n} '(q_{0}), \,{q_{1}
\over q_{0}}J_{n} '(q_{1}), \,{q_{1} \over
q_{0}}I_{n}'(q_{1})\right)
\end{equation}

\begin{equation}
\begin{array}{l}
M_{31} =  -[q_{0}^{2}+(\mu_{0}-2)n^{2}]H_{n}^{(1)}(q_{1})
+q_{0}(\mu_{0}-1)H_{n}^{(1)} {}'(q_{1}) \\
M_{32} = (q_{0}^{2}-\mu_{0}n^{2})K_{n}(q_{0})+
q_{0}(\mu_{0}-1)K_{n} '(q_{0}) \\
M_{33} ={D_{1} \over D_{0}}\left[ (q_{1}^{2}+
(\mu_{1}-2)n^{2})J_{n}(q_{1})
-q_{1}(\mu_{1}-1)J_{n}'(q_{1}) \right]\\
M_{34} = -{D_{1} \over D_{0}}\left[ (q_{1}^{2}-
\mu_{1}n^{2})I_{n}(q_{1}) +q_{1}(\mu_{1}-1)I_{n} '(q_{1})
\right] \\
M_{41} =
-q_{0}(q_{0}^{2}-\mu_{0}n^{2})H_{n}^{(1)} {}'(q_{0})
-n^{2}(\mu_{0}-1)H_{n}^{(1)}(q_{0}) \\
M_{42} = q_{0}[q_{0}^{2} + (\mu_{0}-2)n^{2}]K_{n}'(q_{0})
-n^{2}(\mu_{0}-1)K_{n}(q_{0}) \\
M_{43} = {D_{1} \over D_{0}}\left[
q_{1}(q_{1}^{2}-\mu_{1}n^{2}) J_{n} '(q_{1}) +
(\mu_{1}-1)n^{2}J_{n}(q_{1})\right] \\
M_{44} = -{D_{1} \over D_{0}}\left[q_{1}(q_{1}^{2}
+(\mu_{1}-2)n^{2})I_{n} '(q_{1})-(\mu_{1}-1)n^{2}
I_{n}(q_{1})\right]
\end{array} \label{M}
\end{equation}

\noindent $J_{n} '(x)$, {\it etc.} denote the derivatives of
Bessel functions.



The matrix ${\bf N}$ is identical to ${\bf M}$ except that
the first column, $M_{i1}$, is replaced by the column
vector ${\bf B}$.

\newpage

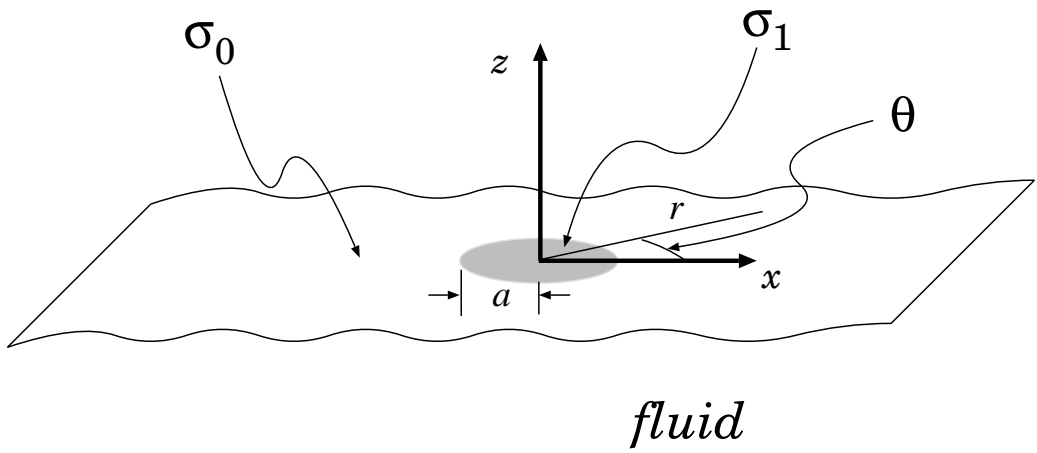
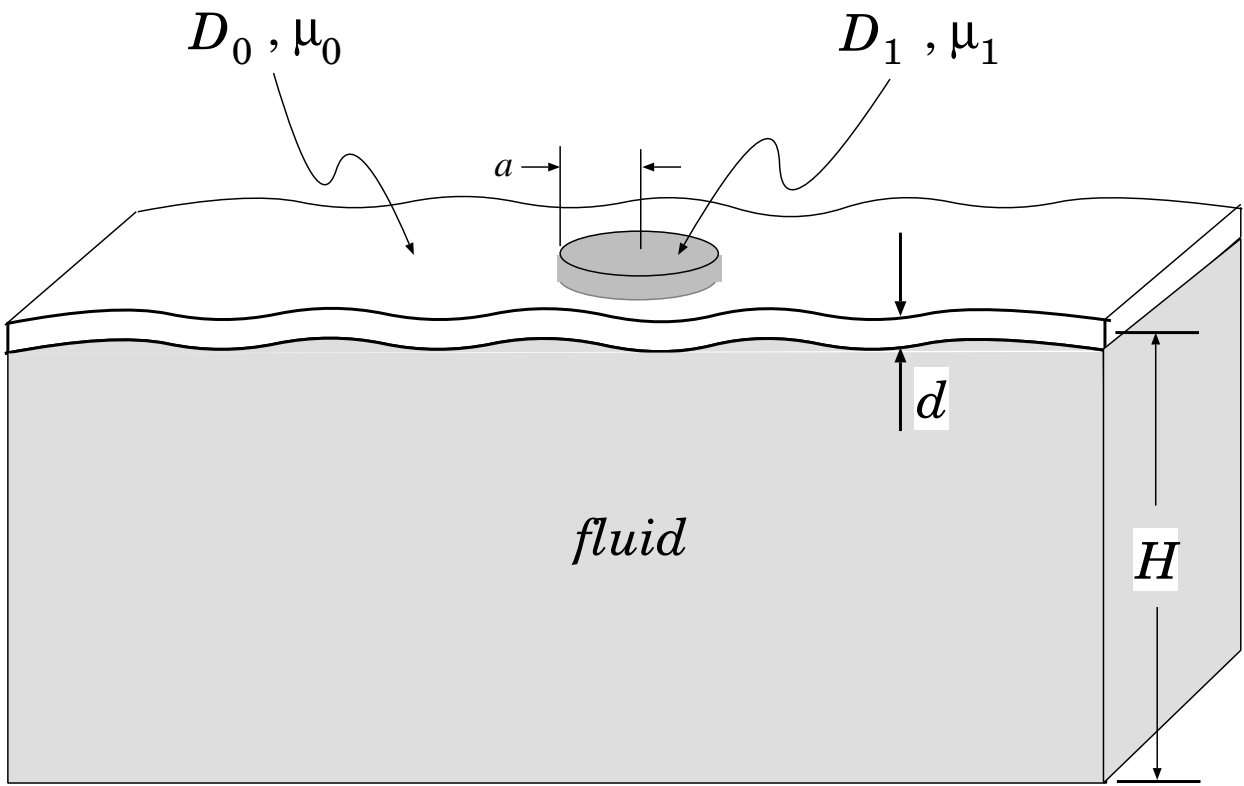
\begin{figure}
\caption{(a) A membrane with a circular domain of
surface tension $\sigma_{1}$ imbedded in an interface of
tension $\sigma_{0}$ .  (b) A plate of thickness $d$ and
flexural rigidity $D_{0}$ with a circular inclusion of
rigidity $D_{1}$.  Ripples in the surface represent the
membrane fluctuations due to an incident plane wave
and the scattered wave.}
\end{figure}

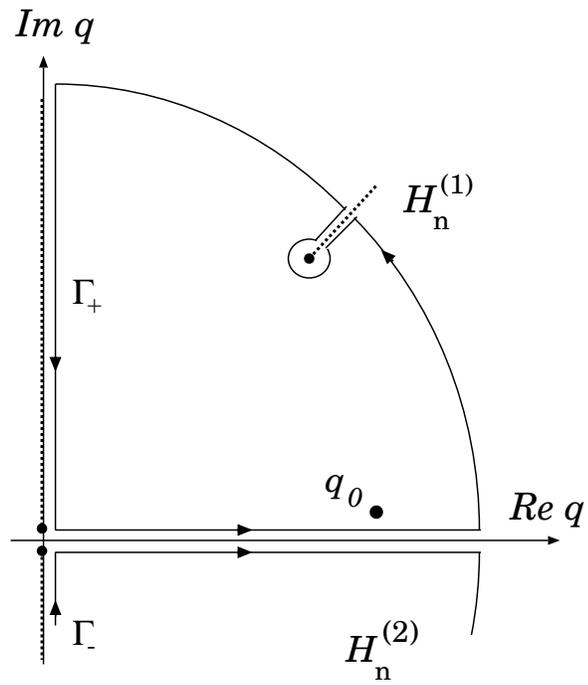
\begin{figure}
\caption{The contour used to evaluate $L_{ml}^{(n)}$ and
$\psi(r,\theta,z)$. An infinitesimally small positive
viscosity and imposition of an outgoing solution requires
that either the pole requires a small positive imaginary
part, or that we take the contour through a semicircle
below the pole.}
\end{figure}

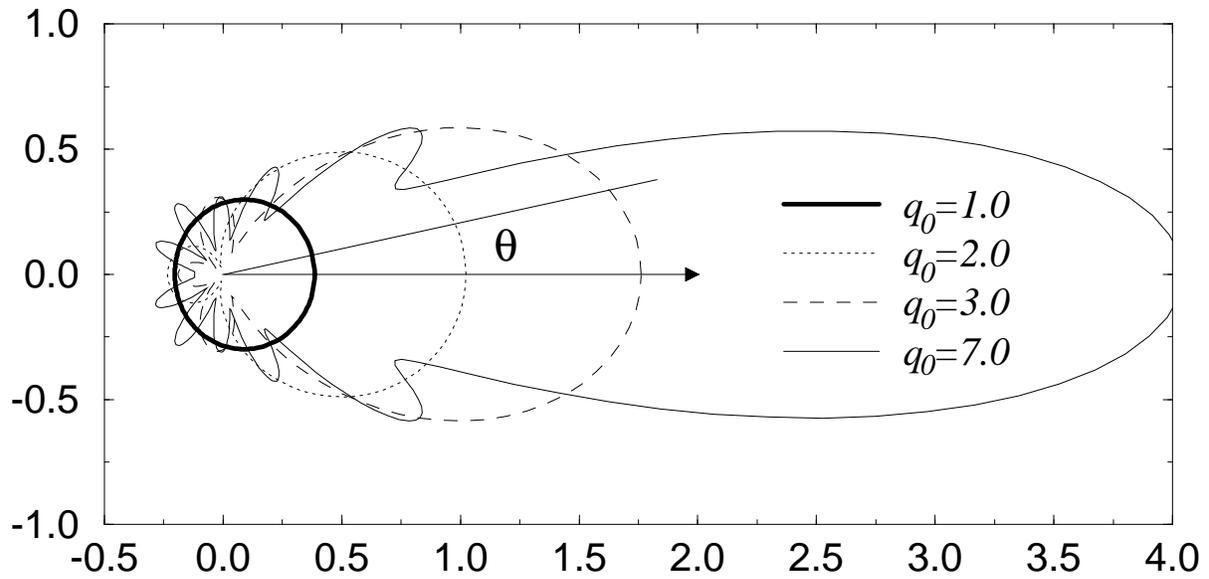
\begin{figure}
\caption{Scattering amplitudes $\vert f(\theta) \vert$ for
an infinite depth system with $\Lambda =
\sigma_{0}/\sigma_{1} = 2.$ The same plots for the
uncoupled membrane, equation (2.50),
at $T_{0}/T_{1} = 2^{2/3}$ are
nearly identical.}
\end{figure}

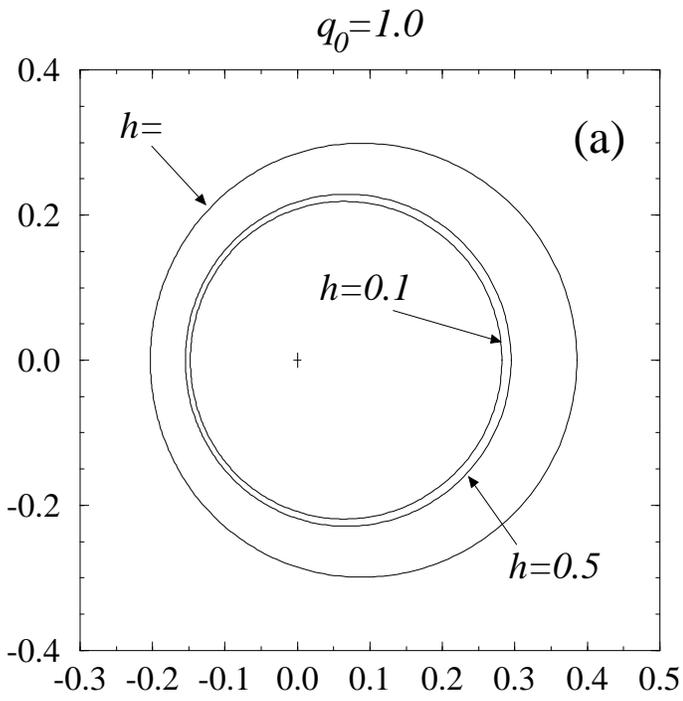
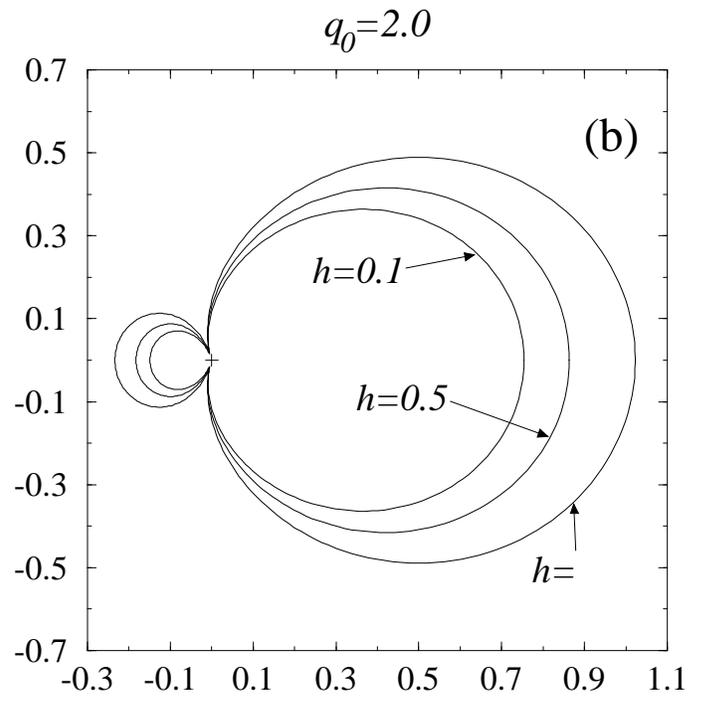
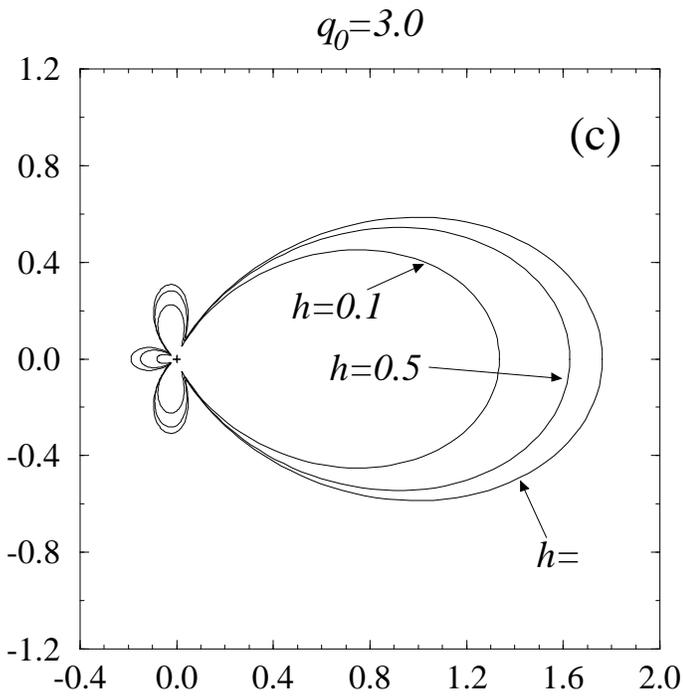
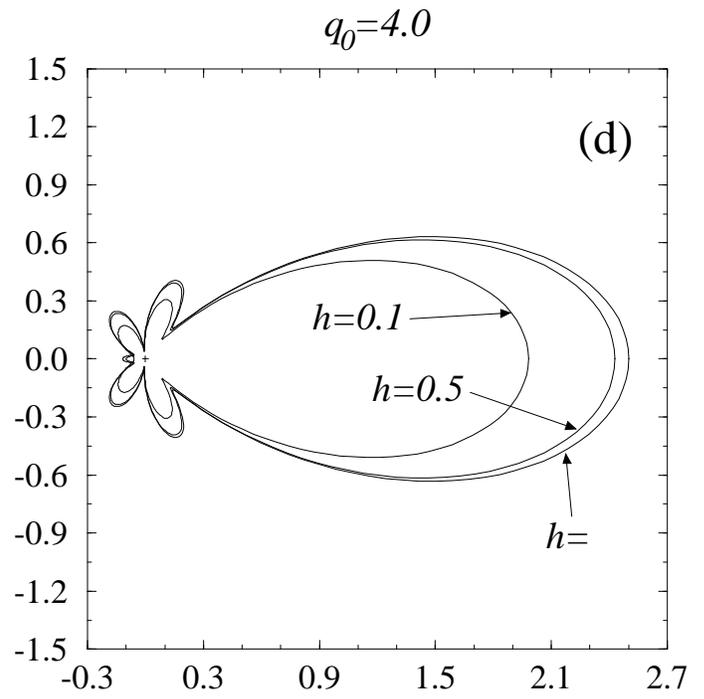
\begin{figure}
\caption{The scattering amplitudes $\vert f(\theta;h)
\vert$ for various depths $h= \infty, 0.5, 0.1$ at $q_{0}=
1,2,3,4$ for $\Lambda=2$. A ``+'' identifies the origin.}
\end{figure}

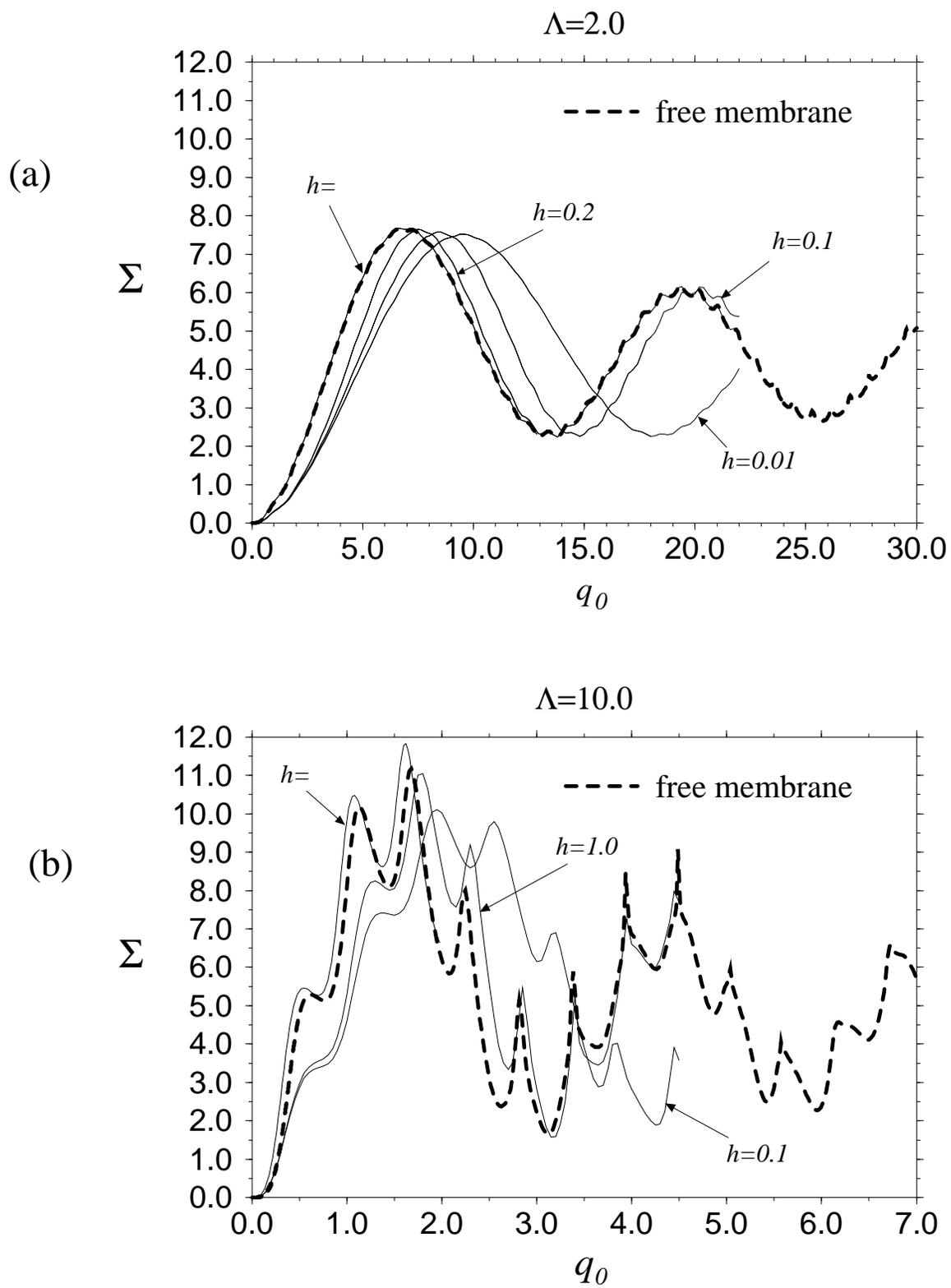
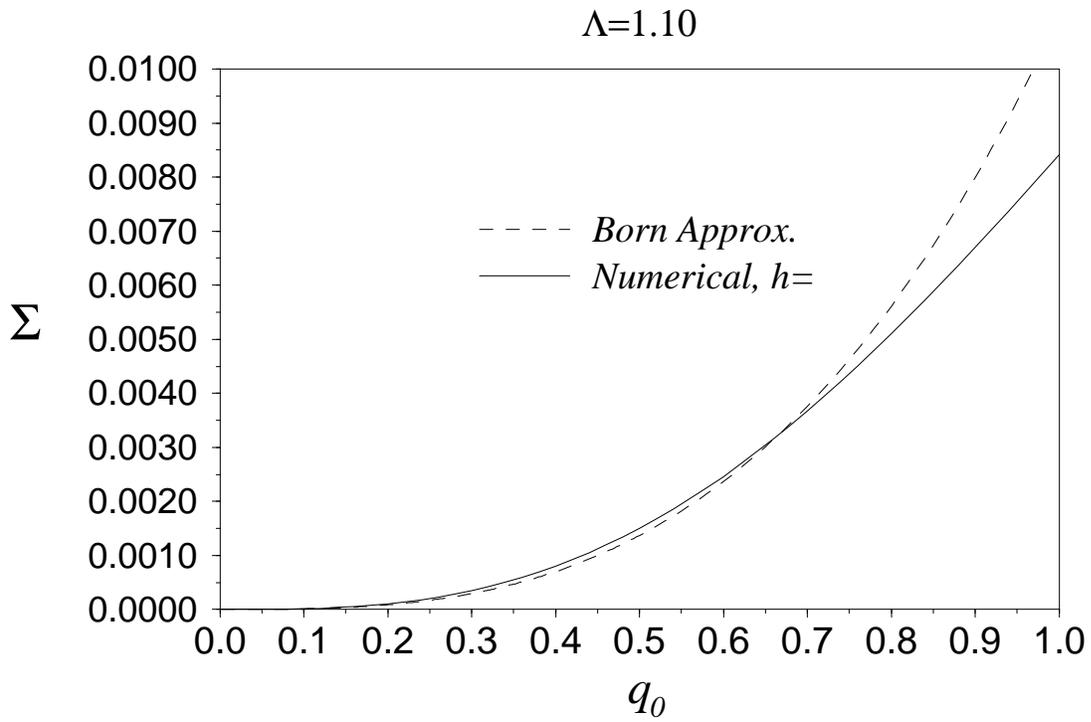
\begin{figure}
\caption{(a) Total scattering cross section
$\Sigma(q_{0},\Lambda=2,h)$ for the surface
tension discontinuity problem. Four depths
are considered: $h=\infty, 0.2,0.1$,
and $0.01.$  (b) Total scattering cross section
$\Sigma(q_{0}, \Lambda=10,h)$ for $h=\infty, 1$, and
$0.1.$  (c) A comparison of the scattering cross section
calculated numerically and asymptotically in the
small $q_0$ limit for $\Lambda= 1.1$. In both (a) and (b)
the dark dashed curve is $\Sigma$ in the free membrane
case, Section IIE.}
\end{figure}

\begin{figure}
\caption{Depth dependence of
$\Sigma(q_{0},\Lambda=2,h)$ for
scattering from a circular surface tension
variation at incoming wavevectors,
$q_{0} = 1.0, 2.0, 3.0, 4.0, 5.0, 8.0, 9.0, $ and $10.0$}
\end{figure}

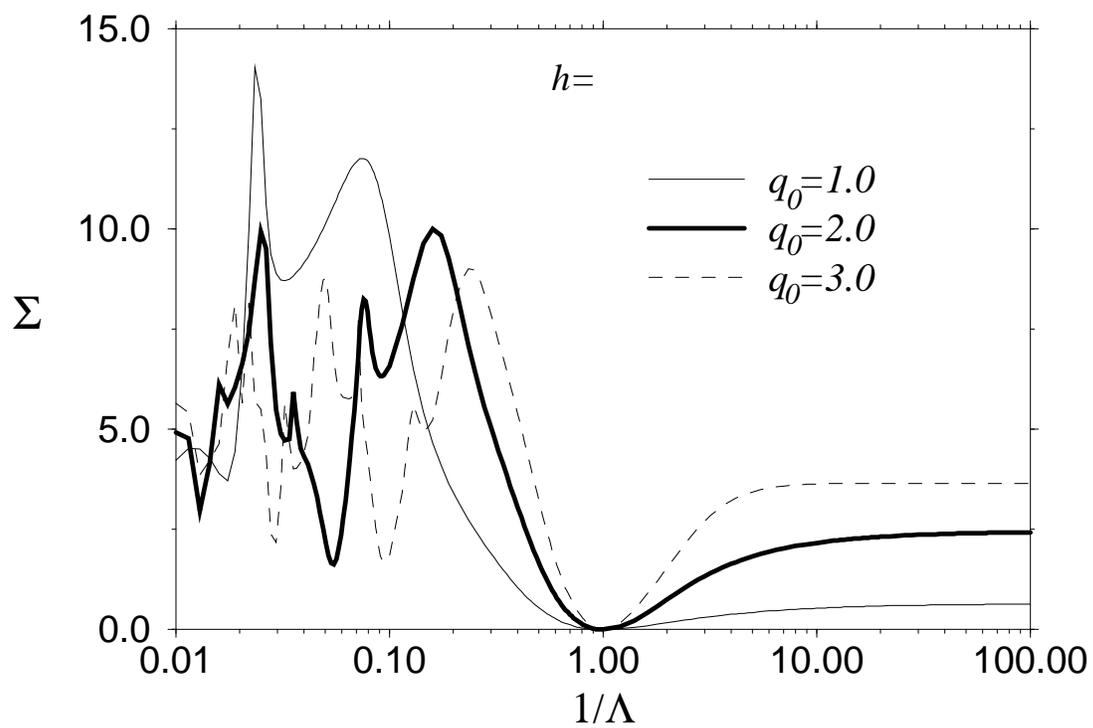
\begin{figure}
\caption{$\Lambda$ dependence of $\Sigma(q_{0},
\Lambda, h=\infty)$ on surface tension ratios. For
$\Lambda^{-1} > 1$, mechanical stability requires an
externally imposed negative line tension. The limit
$\Lambda^{-1} \rightarrow \infty$ corresponds to an
unstretchable domain. }
\end{figure}

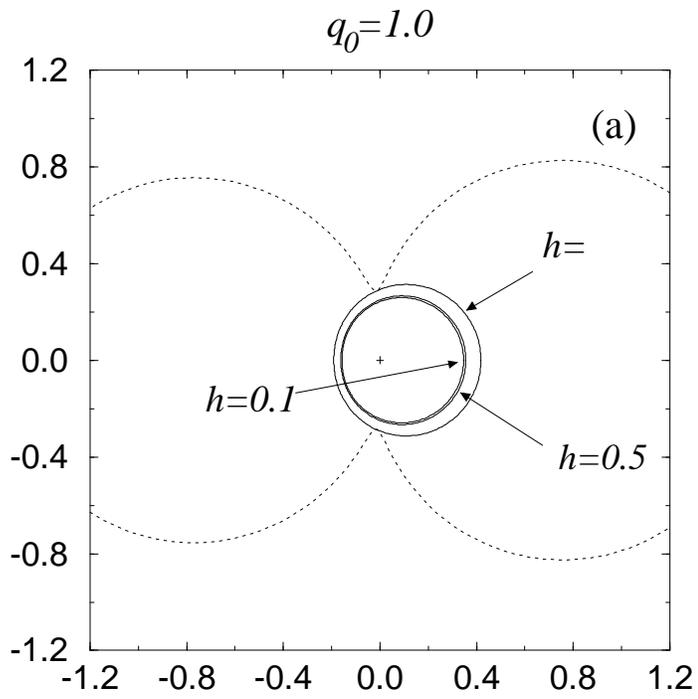
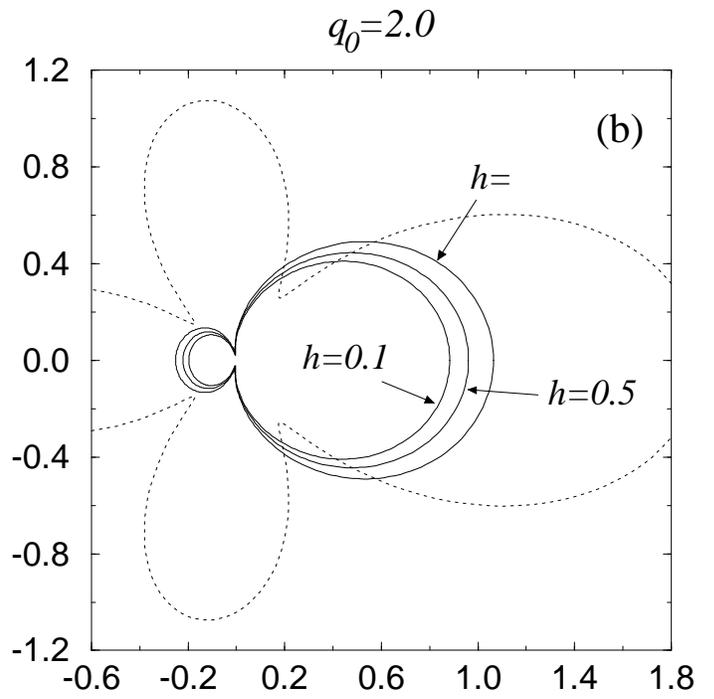
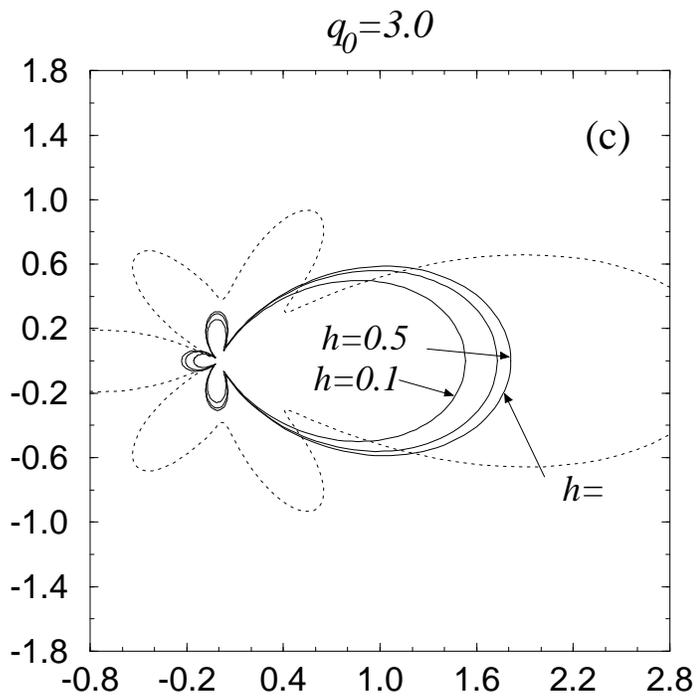
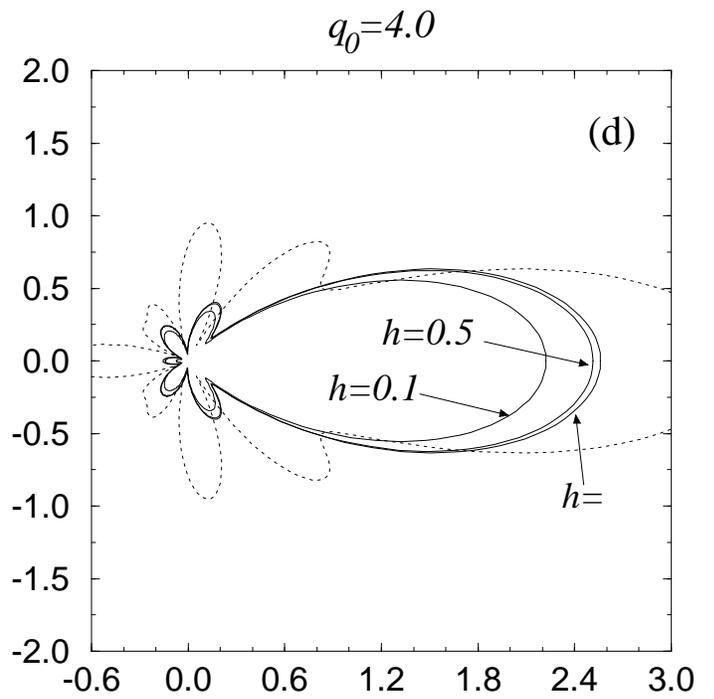
\begin{figure}
\caption{The scattering amplitudes for  an interface with
bending rigidity; $h=\infty, 0.5, 0.1$ and
$\Lambda_{d}=2^{5/3}$.  The dotted curves are
scattering from a free elastic plate. In this case, the
Poisson's ratios are $\mu_{0}=0.25, \mu_{1}=-0.25$. }
\end{figure}

\begin{figure}
\caption{Total cross sections of bending wave scattering
for fluid depths $h=\infty,0.2,0.1,0.01.$ The dotted line is
for a free plate with $\mu_{0}=0.25, \mu_{1}=-0.25$.}
\end{figure}

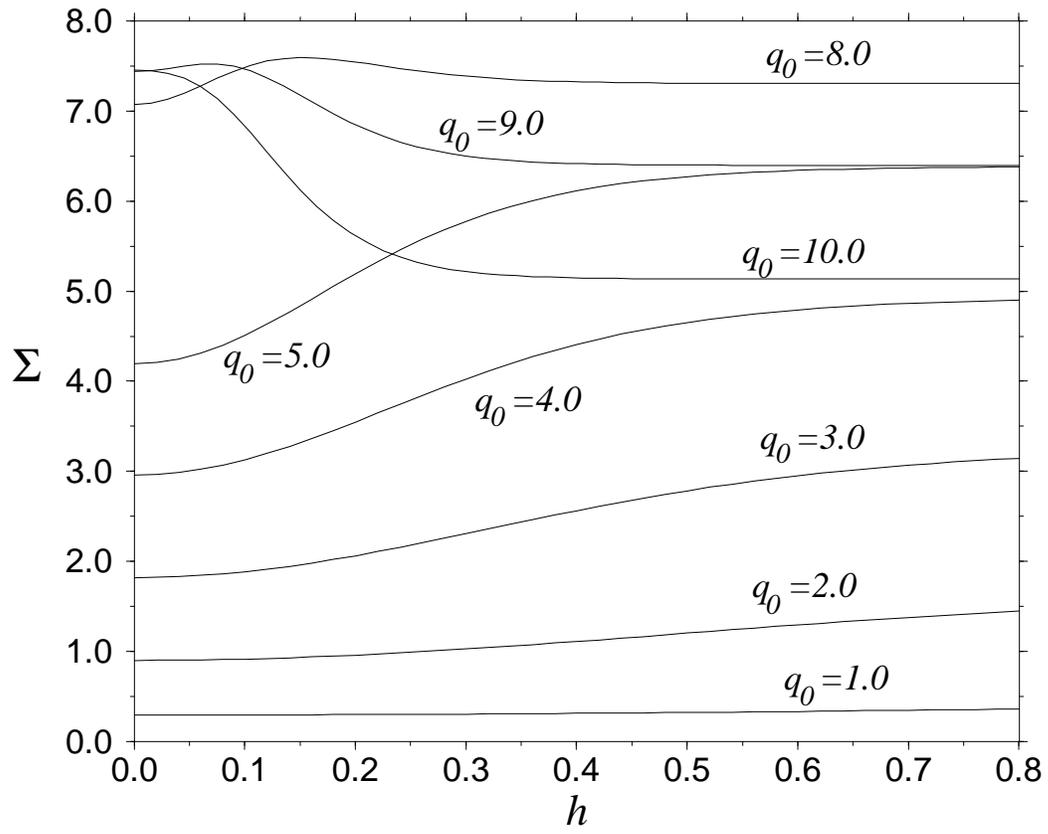
\end{document}